

%
%
%
%

                       \input amstex
\magnification=\magstep1
\baselineskip=12pt
\hfuzz=1pt
\TagsOnRight
\font\eightrm=cmr8
\font\eighti=cmmi8
\font\eightbf=cmbx8
\font\eightit=cmti8

\font\eightsy=cmsy8
\font\sixrm=cmr6
\font\sixi=cmmi6
\font\sixsy=cmsy6

\font\twelverm=cmr10 scaled 1200
\font\twelvebf=cmbx10 scaled 1200
\def\today{\number\day\ \ifcase\month\or
  January\or February\or March\or April\or May\or June\or
  July\or August\or September\or October\or November\or
December\fi \space \number\year}
\def\monthyear{\ifcase\month\or
  January\or February\or March\or April\or May\or June\or
  July\or August\or September\or October\or November\or
December\fi \space \number\year}
\def\eightpoint{\def\rm{\fam0\eightrm}%
  \textfont0=\eightrm \scriptfont0=\sixrm
                      \scriptscriptfont0=\fiverm
  \textfont1=\eighti  \scriptfont1=\sixi
                      \scriptscriptfont1=\fivei
  \textfont2=\eightsy \scriptfont2=\sixsy
                      \scriptscriptfont2=\fivesy
  \textfont3=\tenex   \scriptfont3=\tenex
                      \scriptscriptfont3=\tenex
  \textfont\itfam=\eightit  \def\it{\fam\itfam\eightit}%
  \textfont\bffam=\eightbf  \def\bf{\fam\bffam\eightbf}%
  \normalbaselineskip=16 truept
  \setbox\strutbox=\hbox{\vrule height11pt depth5pt width0pt}}
\tracingstats=1    
\def\CA{{\Cal A}}
\def\hatCA{{\widehat{\Cal A}}}

\def\CF{{\Cal F}}
\def\hatCF{{\widehat\CF}}
\def\hatGamma{{\widehat\Gamma}}
\def\bGamma{{\bar\Gamma}}
\def\SCF{{{\CF}^{(1|1)}}}
\def\SCFg{{{\CF}^{(1|1)}(\gamma)}}
\def\hrSCF{{{\hatCF}^{(1|1)}_{red}}}
\def\hatSCF{{{\hatCF}^{(1|1)}}}
\def\hatSCFg{{{\hatCF}^{(1|1)}(\gamma)}}
\def\rSCF{{{\CF}^{(1|1)}_{red}}}
\def\hySCF{{{\hatCF}^{(1|1)}_{y_M}}}

\def\CH{{\Cal H}}
\def\SCH{{{\Cal H}^{(1|1)}}}

\def\CI{{\Cal I}}
\def\CL{{\Cal L}}
\def\CM{{\Cal M}}

\def\widehatC{\widehat{\bbbc}}
\def\bbbr{\operatorname{{I\!R}}} 
\def\bbbn{\operatorname{{I\!N}}} 

\def\bbbone{\mathchoice {\operatorname{1\mskip-4mu l}}
{\operatorname{1\mskip-4mu l}}
{\operatorname{1\mskip-4.5mu l}} {\operatorname{1\mskip-5mu l}}}
\def\bbbc{{\mathchoice {\setbox0=\hbox{\rm C}\hbox{\hbox
to0pt{\kern0.4\wd0\vrule height0.9\ht0\hss}\box0}}
{\setbox0=\hbox{$\textstyle\hbox{\rm C}$}\hbox{\hbox
to0pt{\kern0.4\wd0\vrule height0.9\ht0\hss}\box0}}
{\setbox0=\hbox{$\scriptstyle\hbox{\rm C}$}\hbox{\hbox
to0pt{\kern0.4\wd0\vrule height0.9\ht0\hss}\box0}}
{\setbox0=\hbox{$\scriptscriptstyle\hbox{\rm C}$}\hbox{\hbox
to0pt{\kern0.4\wd0\vrule height0.9\ht0\hss}\box0}}}}
\def\bbbe{{\mathchoice {\setbox0=\hbox{\smalletextfont e}\hbox{\raise
0.1\ht0\hbox to0pt{\kern0.4\wd0\vrule width0.3pt
height0.7\ht0\hss}\box0}}
{\setbox0=\hbox{\smalletextfont e}\hbox{\raise
0.1\ht0\hbox to0pt{\kern0.4\wd0\vrule width0.3pt
height0.7\ht0\hss}\box0}}
{\setbox0=\hbox{\smallescriptfont e}\hbox{\raise
0.1\ht0\hbox to0pt{\kern0.5\wd0\vrule width0.2pt
height0.7\ht0\hss}\box0}}
{\setbox0=\hbox{\smallescriptscriptfont e}\hbox{\raise
0.1\ht0\hbox to0pt{\kern0.4\wd0\vrule width0.2pt
height0.7\ht0\hss}\box0}}}}
\def\bbbq{{\mathchoice {\setbox0=\hbox{$\displaystyle\rm Q$}\hbox{\raise
0.15\ht0\hbox to0pt{\kern0.4\wd0\vrule height0.8\ht0\hss}\box0}}
{\setbox0=\hbox{$\textstyle\rm Q$}\hbox{\raise
0.15\ht0\hbox to0pt{\kern0.4\wd0\vrule height0.8\ht0\hss}\box0}}
{\setbox0=\hbox{$\scriptstyle\rm Q$}\hbox{\raise
0.15\ht0\hbox to0pt{\kern0.4\wd0\vrule height0.7\ht0\hss}\box0}}
{\setbox0=\hbox{$\scriptscriptstyle\rm Q$}\hbox{\raise
0.15\ht0\hbox to0pt{\kern0.4\wd0\vrule height0.7\ht0\hss}\box0}}}}
\def\bbbs{{\mathchoice
{\setbox0=\hbox{$\displaystyle     \rm S$}\hbox{\raise0.5\ht0\hbox
to0pt{\kern0.35\wd0\vrule height0.45\ht0\hss}\hbox
to0pt{\kern0.55\wd0\vrule height0.5\ht0\hss}\box0}}
{\setbox0=\hbox{$\textstyle        \rm S$}\hbox{\raise0.5\ht0\hbox
to0pt{\kern0.35\wd0\vrule height0.45\ht0\hss}\hbox
to0pt{\kern0.55\wd0\vrule height0.5\ht0\hss}\box0}}
{\setbox0=\hbox{$\scriptstyle      \rm S$}\hbox{\raise0.5\ht0\hbox
to0pt{\kern0.35\wd0\vrule height0.45\ht0\hss}\raise0.05\ht0\hbox
to0pt{\kern0.5\wd0\vrule height0.45\ht0\hss}\box0}}
{\setbox0=\hbox{$\scriptscriptstyle\rm S$}\hbox{\raise0.5\ht0\hbox
to0pt{\kern0.4\wd0\vrule height0.45\ht0\hss}\raise0.05\ht0\hbox
to0pt{\kern0.55\wd0\vrule height0.45\ht0\hss}\box0}}}}
\def\bbbt{{\mathchoice {\setbox0=\hbox{$\displaystyle\rm
T$}\hbox{\hbox to0pt{\kern0.3\wd0\vrule height0.9\ht0\hss}\box0}}
{\setbox0=\hbox{$\textstyle\rm T$}\hbox{\hbox
to0pt{\kern0.3\wd0\vrule height0.9\ht0\hss}\box0}}
{\setbox0=\hbox{$\scriptstyle\rm T$}\hbox{\hbox
to0pt{\kern0.3\wd0\vrule height0.9\ht0\hss}\box0}}
{\setbox0=\hbox{$\scriptscriptstyle\rm T$}\hbox{\hbox
to0pt{\kern0.3\wd0\vrule height0.9\ht0\hss}\box0}}}}
\font\sans=cmssbx10
\def\sf{\sans}
\def\bbbz{{\mathchoice {\hbox{$\sf\textstyle Z\kern-0.4em Z$}}
{\hbox{$\sf\textstyle Z\kern-0.4em Z$}}
{\hbox{$\sf\scriptstyle Z\kern-0.3em Z$}}
{\hbox{$\sf\scriptscriptstyle Z\kern-0.2em Z$}}}}

\def\Cup{\bigcup}

\def\Li{{\Lambda_\infty}}
\def\viert{{1\over4}}
\def\half{{1\over2}}
\def\bhalf{\hbox{$\half$}}
\def\tkappa{\kappa_0}
\def\mZO{{{\hat Z}_0}}
\def\mZEN{{{\hat Z}_0^{(N)}}}
\def\mZE{{{\hat Z}_1}}
\def\mZEz{{{\hat Z}_1^2}}
\def\mRO{{{\hat R}_0}}
\def\mROp{{{\hat R}_0^{\prime}}}
\def\mRE{{{\hat R}_1}}
\def\mREp{{{\hat R}_1^{\prime}}}
\def\mZS{{{\hat Z}_S}}
\def\mZSp{{{\hat Z}_S^{\prime}}}

\def\mPsiE{{{\hat\Psi}_1}}
\def\mPsiO{{{\hat\Psi}_0}}
\def\mPsiS{{{\hat\Psi}_S}}

\def\Aut{\operatorname{Aut}}

\def\dim{\operatorname{dim}}
\def\ker{\operatorname{ker}}
\def\mod{\operatorname{mod}}
\def\PSL{\operatorname{PSL}}
\def\GL{\operatorname{GL}}
\def\SL{\operatorname{SL}}
\def\order{\operatorname{order}}
\def\OSp{\operatorname{OSp}}
\def\sdet{\operatorname{sdet}}
\def\sign{\operatorname{sign}}
\def\str{\operatorname{str}}
\def\tr{\operatorname{tr}}

\def\sqr#1#2{{\vcenter{\hrule height.#2pt
              \hbox{\vrule width.#2pt height#1pt
              \kern#1pt
              \vrule width.#2pt}
              \hrule height.#2pt}}}
\def\square{\mathchoice{\sqr68}{\sqr68}{\sqr{2.1}3}{\sqr{1.5}3}}
\def\hsquare{\widehat{
              \mathchoice{\sqr68}{\sqr68}{\sqr{2.1}3}{\sqr{1.5}3}}}

\hyphenation{Vols de-ve-lop-ed Rie-mann Nucl the-ory straight-for-ward}

\newcount\glno
\def\plus{\advance\glno by 1}
\def\num{\the\glno}

\newcount\chapno
\def\NUM{\the\chapno}

\newcount\refno
\def\add{\advance\refno by 1}
\refno=1

\edef\ALV{\the\refno}\add
\edef\ALVA{\the\refno}\add
\edef\AOK{\the\refno}\add
\edef\BMFSa{\the\refno}\add
\edef\BMFSb{\the\refno}\add
\edef\BFS{\the\refno}\add
\edef\BASCH{\the\refno}\add
\edef\BATCH{\the\refno}\add
\edef\BABR{\the\refno}\add
\edef\BLCL{\the\refno}\add
\edef\BCDPC{\the\refno}\add
\edef\BOGRO{\the\refno}\add
\edef\BOSTa{\the\refno}\add
\edef\BOSTb{\the\refno}\add
\edef\BUMO{\the\refno}\add
\edef\CARL{\the\refno}\add
\edef\DEBRO{\the\refno}\add
\edef\DEW{\the\refno}\add
\edef\DHPHa{\the\refno}\add
\edef\DHPHb{\the\refno}\add
\edef\DHPHc{\the\refno}\add
\edef\DHPHd{\the\refno}\add
\edef\DUN{\the\refno}\add
\edef\EFR{\the\refno}\add
\edef\ELS{\the\refno}\add
\edef\EMOTa{\the\refno}\add
\edef\GIL{\the\refno}\add
\edef\GRA{\the\refno}\add
\edef\GRE{\the\refno}\add
\edef\GSa{\the\refno}\add
\edef\GSb{\the\refno}\add
\edef\GSW{\the\refno}\add
\edef\GROd{\the\refno}\add
\edef\GROe{\the\refno}\add
\edef\GROf{\the\refno}\add
\edef\GP{\the\refno}\add
\edef\HEJ{\the\refno}\add
\edef\HEJa{\the\refno}\add
\edef\HEJb{\the\refno}\add
\edef\HOWE{\the\refno}\add
\edef\JASK{\the\refno}\add
\edef\KOYA{\the\refno}\add
\edef\LOSEV{\the\refno}\add
\edef\LUCK{\the\refno}\add
\edef\MANIN{\the\refno}\add
\edef\MDM{\the\refno}\add
\edef\MUYa{\the\refno}\add
\edef\MUYb{\the\refno}\add
\edef\MNP{\the\refno}\add
\edef\MORO{\the\refno}\add
\edef\NARA{\the\refno}\add
\edef\NINN{\the\refno}\add
\edef\OHN{\the\refno}\add
\edef\OSH{\the\refno}\add
\edef\PHSAR{\the\refno}\add
\edef\POLa{\the\refno}\add
\edef\POLb{\the\refno}\add
\edef\RC{\the\refno}\add
\edef\RODR{\the\refno}\add
\edef\ROVT{\the\refno}\add
\edef\ROG{\the\refno}\add
\edef\SCHW{\the\refno}\add
\edef\SELB{\the\refno}\add
\edef\SHOK{\the\refno}\add
\edef\SIBN{\the\refno}\add
\edef\STEI{\the\refno}\add
\edef\TAZO{\the\refno}\add
\edef\UEYA{\the\refno}\add
\edef\VENa{\the\refno}\add
\edef\VENb{\the\refno}\add
\edef\VENc{\the\refno}\add
\edef\VENd{\the\refno}\add
\edef\VOROS{\the\refno}\add
\edef\WU{\the\refno}\add


{\nopagenumbers
\pageno=0
\centerline{October 1992\hfill SISSA/180/92/FM}
\vskip1cm
\centerline{\twelvebf Selberg Supertrace Formula
                      for Super Riemann Surfaces III:}
\bigskip
\centerline{\twelvebf Bordered Super Riemann Surfaces}
\bigskip
\centerline{by}
\bigskip
\centerline{\twelverm Christian Grosche}
\bigskip
\centerline{\it Scuola Internazionale Superiore di Studi Avanzati}
\centerline{\it International School for Advanced Studies}
\centerline{\it Via Beirut 4}
\centerline{\it 34014 Trieste, Miramare, Italy}
\vfill
\midinsert\narrower
\noindent {\bf Abstract}

\noindent
This paper is the third in a sequel to develop a super-analogue of the
classical Selberg trace formula, the Selberg supertrace formula. It
deals with bordered super Riemann surfaces. The theory of bordered
super Riemann surfaces is outlined, and the corresponding Selberg
supertrace formula is developed. The analytic properties of the Selberg
super zeta-functions on bordered super Riemann surfaces are discussed,
and super-determinants of Dirac-Laplace operators on bordered super
Riemann surfaces are calculated in terms of Selberg super
zeta-functions.
\endinsert
\eject
\pageno=0
\centerline{\ }
\newpage}

\baselineskip=11pt
\pageno=1
\glno=0                
\advance\chapno by 1   
\line{\bf I.\ Introduction\hfill}
\par\medskip\nobreak\noindent
It took a long time before physicists acknowledged the true value of the
Selberg trace formula as introduced by A.Selberg in his famous paper
[\SELB]. The original attempt of Selberg to formulate his trace formula
was based on number theoretical considerations, and in fact, there is a
close relation between the areas of analytic number theory, eigenvalues
of Laplacians on Riemann surfaces, and the Selberg trace formula (see
e.g.\ [\HEJ, \SHOK]), and in particular Selberg was interested to study
the analytic properties of a function closely related to the trace
formula, the Selberg zeta-function.

Physicists, however, have other objectives: they want to learn something
about the spectrum of a model, or they want to calculate determinants,
say. The latter approach to the use of the Selberg trace formula
appears in the quantum field theory on Riemann surfaces, i.e.\ in the
Polyakov approach [\DHPHa-\DHPHd, \POLa, \POLb] to (bosonic-, fermionic-
and super-) string theory. In the perturbation expansion of the Polyakov
path integral one is left with a summation over all topologies of
world sheets a string can sweep out, and an integral over the moduli
space of Riemann surfaces. This picture is true for bosonic strings
(BS) as well as for fermionic strings (FS). The partition function
turns out to be for open as well as closed bosonic strings
corresponding to a topology without conformal Killing vectors (Blau and
Clements [\BLCL], and D'Hoker and Phong [\DHPHa, \DHPHd])
\plus
$$Z_0^{(BS)}=\sum_g\int d\mu_{WP}[\det(P_1^{\dag}P_1)]^{1/2}
           ({\det}'\Delta_0)^{-D/2}.
  \tag\NUM.\num$$
$P_1$ and $\Delta_0$ are the symmetrized traceless covariant derivative
and scalar Laplacian with Dirichlet boundary-conditions, respectively,
and $d\mu_{WP}$ denotes the Weil-Petersen measure. $D$ denotes the
critical dimension which equals $26$ for the bosonic string. For the
fermionic-, respectively the super-string all quantities have to be
replaced by their appropriate super case, and the critical dimension
$D=10$.

The calculation of (super-) determinants of Laplacians on closed Riemann
surfaces is due to several authors. Mainly two approaches must be
mentioned, firstly the evaluation of these determinants in terms of
Selberg zeta-functions, e.g.\ Baranov, Manin et al.\ [\BMFSa-\BASCH],
Bolte and Steiner [\BOSTa], D'Hoker and Phong [\DHPHb, \DHPHd], Efrat
[\EFR], Ref.[\GROd], Gilbert [\GIL], Namazie and Rajjev [\NARA],
Steiner [\STEI], and Voros [\VOROS], and secondly in terms of the
period matrix and theta-functions, the most important are
Alvarez-Gaum\'e et al.\ [\ALVA] and Manin [\MANIN]. Formal as these
results may be, the expressions in terms of Selberg zeta-functions
provide tools to investigate the convergence, respectively divergence
properties of the string path integral a l\`a Polyakov, hence
non-perturbative statements are possible. In the bosonic string theory,
this approach enabled Gross and Periwal [\GP] to show that the
perturbation expansion for the closed bosonic string is not
Borel-summable, and hence not finite; this statement can be easily
generalized to open bosonic strings [\BOGRO]. In the fermionic string
theory a better asymptotic behaviour is expected, however, due to the
almost unknown structure of the corresponding super moduli space, the
arguing of Gross and Periwal cannot be taken over in an obvious way.

In the perturbation expansion of the bosonic string the classical
Selberg trace formula could be applied in a straightforward way; the
perturbation theory of the fermionic string [\DHPHc, \DHPHd] required
the devolving of a super-analogue of the classical Selberg  trace
formula, i.e.\ the Selberg supertrace formula. Here Baranov, Manin et
al.\ [\BMFSa-\BASCH] originally started this business, and it was
further developed by Aoki [\AOK] and in Refs.[\GROd, \GROe], and see
[\GROf] for a short review of some recent results.

It was mainly the closed string theory that was dealt with and for which
the whole perturbation theory for scattering amplitudes was developed
quite comprehensively. The case of the open bosonic, respectively open
fermionic string, starting with the pioneering work of Alvarez [\ALV],
took somewhat longer and seems until now not such as good developed as
the former.

Contributions along the lines of the Polyakov path integral approach for
the open string theory are due to e.g.\ Blau et al.\ [\BLCL, \BCDPC],
Bolte and Grosche [\BOGRO], Bolte and Steiner [\BOSTb], Burgess and
Morris [\BUMO], Carlip [\CARL], Dunbar [\DUN], Jaskolski [\JASK],
Luckock [\LUCK], Mart\'\i n-Delgado and Mittelbrunn [\MDM], Ohndorf
[\OHN], Rodrigues et al.\ [\DEBRO, \RODR, \ROVT], and Wu [\WU].

Of course, while dealing with open strings one has to distinguish
Dirichlet and Neumann boundary-conditions, respectively. Here again we
have two possibilities to express determinants either by the
period-matrix, or by appropriate chosen  Selberg zeta-functions for the
corresponding Dirichlet or Neumann boundary-condition problems. The
former was discussed by Burgess and Morris [\BUMO], Dunbar [\DUN],
Losev [\LOSEV], Luckock [\LUCK], Mart\'\i n-Delgado and Mittelbrunn
[\MDM], Rodrigues and Van Tander [\ROVT] (in particular to give
explicit expression on the one-loop level), and Mozorov and Rosly
[\MORO] for multiloop expressions. The latter case was approached by
Refs.[\BLCL-\BOGRO, \BOSTb].

In particular all the cited authors could derive relations between the
determinants $\det\Delta^{(D)}_{\Sigma}$ and ${\det}'\Delta^{(N)}
_{\Sigma}$ corresponding to Dirichlet and Neumann boundary-conditions
on the bordered Riemann surface $\Sigma$ on the one hand, and the
determinant of the scalar Laplacian ${\det}'\Delta_{\widehat\Sigma}$
for the doubled (closed) Riemann surfaces $\widehat\Sigma$ on the other,
i.e.\ (the prime denotes the omission of possible zero-modes)
\plus
$$\det\Delta^{(D)}_{\Sigma}\cdot{\det}'\Delta^{(N)}_{\Sigma}
  ={\det}'\Delta_{\widehat\Sigma}.
  \tag\NUM.\num$$
\edef\numab{\NUM.\num}%
Whereas all this deals only with the bosonic string with boundaries, the
case of the incorporation of spin-structures and fermionic strings with
boundaries seems quite poorly developed.

The Selberg trace formula for bordered Riemann surfaces does exist and
almost the entire theory is due to Venkov [\VENb-\VENd], including
particular cases [\VENa]. Independently, later on the Selberg trace
formula for bordered Riemann surfaces was discussed by Blau and
Clements [\BLCL], Bolte and Grosche [\BOGRO] and Bolte and Steiner
[\BOSTb].

Let us assume that the generating functional in the theory of the open
fermionic string can also be expressed as [\DHPHc, \DHPHd]
\plus
$$Z_0^{(FS)}=\sum_g\int d\mu_{sWP}[\sdet(P_1^{\dag}P_1)]^{1/2}
                          (\sdet'\square^2_0)^{-5/2},
  \tag\NUM.\num$$
\edef\numac{\NUM.\num}%
where $P_1$ and $\square_0$ are the super-analogues of the symmetrized
traceless covariant derivative and scalar Dirac-Laplace operator with
Dirichlet boundary-conditions, respectively, and $d\mu_{sWP}$ denotes
the super Weil-Petersen measure. In order to deal with the vector
Dirac-Laplace operator $P_1^{\dag}P_1$ the incorporation of
$m$-weighted super automorphic forms into the formalism is required.

\eject
The development of the fermionic string, respectively the super-string
model [\GRE-\GSW,\SCHW] was enormously boosted by the discovery of
particular anomaly-free properties of certain gauge groups by Green and
Schwarz [\GSb]. In two previous publications, hereafter denoted as I
[\GROd], and II [\GROe], respectively, I have already discussed various
aspects of the Selberg supertrace formula on super Riemann surfaces.
In Ref.[\GROd] the Selberg supertrace formula for hyperbolic
conjugacy classes was developed in full detail, including an analysis
of the properties of the Selberg super zeta-functions and
super-determinants of Dirac-Laplace operators on Super Riemann surfaces
in order to discuss the fermionic string integrand in the Polyakov path
integral properly. In Ref.[\GROe], I continued these studies by the
incorporation of elliptic and parabolic conjugacy classes. However, to
complete a comprehensive development of the Selberg supertrace formula,
bordered super Riemann surfaces must be included in the discussion.
Bordered super Riemann surfaces, of course, occur for open (fermionic-
and super-) strings (in the case of super-strings, so called type I
super strings with the $O(32)$ gauge group).

The further contents will be now as follows:

In the second section I give a short introduction into the theory of
bordered Riemann surfaces and indispensible information concerning the
spectral theory on bordered Riemann surfaces. Information concerning the
definition of super Riemann surfaces was already given in Refs.[\GROd,
\GROe], but is included here to make the paper self-contained. The
citation of the Selberg trace formula on bordered Riemann surfaces is
also included.

The third section sets up a proposal to define bordered super
Riemann surfaces. Concerning the theory of super Riemann surfaces I
refer to the relevant literature, only the most important relations and
indispensible ingredients will be given.

The fourth section attacks the actual derivation of the Selberg
supertrace formula for bordered super Riemann surfaces. After the
incorporation of the hyperbolic conjugacy classes, the usual elliptic
ones cause no difficulty and the result of Ref.[\GROe] can be taken
over without delay; the incorporation of the parabolic conjugacy
classes (Dirichlet boundary-conditions) requires some care and a
regularization procedure is needed. The incorporation of Neumann
boundary-conditions then follows by a proper combination of the former
results.

In the fifth section, I discuss the analytic properties of the
(modified) Selberg super zeta-functions on bordered super Riemann
surfaces by means of the supertrace formula.

In the sixth section, the results of the former two are applied to the
problem of calculating super-determinants of Dirac-Laplace operators on
bordered super Riemann surfaces. As explicit as possible expressions are
evaluated in terms of the Selberg super zeta-functions on bordered super
Riemann surfaces.

All the principle results will be stated as theorems.

The last section is devoted to a summary and a discussion of the
results.


\bigskip\goodbreak\noindent
\glno=0                
\advance\chapno by 1   
\line{\bf II.\ The Selberg Trace Formula on Bordered Riemann
      Surfaces\hfill}
\par\medskip\nobreak\noindent
{\it 1.\ Bordered Riemann surfaces and the construction of the Selberg
operator.}
We start with usual bordered Riemann surfaces (we rely on Sibner [\SIBN]
and Venkov [\VENc], compare also Refs.[\BOGRO, \BOSTb]). Let us
consider a Riemann $\widetilde\Sigma$ surface of genus $g$ and
$d_1,\dots,d_n$ conformal, non-overlap\-ping discs on $\widetilde
\Sigma$. Then $\Sigma=\widetilde\Sigma\setminus\{d_1,\dots,d_n\}$ is a
bordered Riemann surface with Dirichlet, respectively Neumann
boundary-conditions of signature $(g,n)$. $c_i=\partial d_i$ are the
$n$ components of $\partial\Sigma$. Now take a copy $\CI\Sigma$ of
$\Sigma$, a mirror image, and glue both together along $\partial\Sigma$
and $\partial\CI\Sigma$. Explicitly this can be realized by taking the
refection $\CI$ to have the form $\CI z\to z'=-\bar z$ $(z\in\CH)$. The
reflection $\CI$ in $\partial\Sigma$ then is a anti-conformal
involution ($\CI^2=1)$ on the doubled surface $\widehat\Sigma
=\Sigma\cup\CI\Sigma$. $\widehat\Sigma$ is a Riemann surface of genus
$\widehat g=2g+n-1$, and we set $\hatCA=\CA(\hatCF)$. The
uniformization theorem for Riemann surfaces now yields that $\widehat
\Sigma$ may be represented as $\widehat\Sigma\cong\hatGamma
\backslash\CH$, where $\hatGamma$ is the Fuchsian group of
$\widehat\Sigma$. Of course there is a fundamental domain $\widehat\CF$
of $\widehat\Sigma$ in $\CH$, tesselating $\CH$. In order to construct
a convenient fundamental domain and representation of the involution
$\CI$, we view according to Sibner [\SIBN] and Venkov [\VENc]
$\widehat\Sigma$ as a symmetric Riemann surface with reflection symmetry
$\CI$. Then $\widehat\CF$ may be chosen as the interior of a
fundamental polygon with $4\widehat g+2n-2$ edges which is symmetric
with respect to the imaginary axis. Due to the explicit choice of $\CI$
as $\CI z=-\bar z$, one of the bordering curves, say $c_1$, is mapped
onto the imaginary axis and the others among the edges of the
fundamental polygon. With this construction one can work directly on
$\hatCF$, with $\CI$ viewed as a mapping of complex numbers, being
formally identical on $\widehat\Sigma$ and $\hatCF$.

In the case that elliptic fixed points and cusps are present, the
non-euclidean area of a Riemann surface is given by (e.g.\ [\HEJb],
Gauss-Bonnet theorem)
\plus
$$\CA=2\pi\left[2(g-1)+\kappa+\sum_{j=1}^s
    \bigg(1-{1\over\nu_j}\bigg)\right],
  \tag\NUM.\num$$
where $s$ denotes the number of inequivalent elliptic fixed points and
$\kappa$ the number of inequivalent cusps (i.e.\ the number of zero
interior angles of the fundamental polygon $\CF$). $\nu_j$ denotes the
order of the elliptic generators $R_j\subset\Gamma$ ($1\leq j\leq s$),
i.e.\ $R_j^{\nu_j}=1$ for $(1\leq j\leq s$, $1<\nu_j<\infty$).

In order to set up our notation we start by citing some results of the
classical Selberg trace formula for bordered Riemann surfaces [\VENc].
As usual one starts by formulating the appropriate automorphic kernel.
Consequently this gives for automorphic functions the property
$f(\gamma z)=\chi_\gamma^mj(\gamma,z)f(z)$ $(\gamma\in\hatGamma)$,
the inversion $f(\CI z)=\pm f(z)$ which distinguishes even and odd
automorphic functions with respect to $x$. $\chi(\gamma)\equiv
\chi_\gamma$ denotes a multiplier system acting according to
$\bGamma\to \{\pm1\}$ with $\chi(-\bbbone)=-1$ ($\gamma\in\bGamma$,
such that $\bGamma\in\SL(2,\bbbr)$, $\hatGamma=\bGamma/\{\pm1\}$), and
$j(\gamma,z)$ denotes the automorphic weight [\BOGRO, \HEJa].

The automorphic kernel is then constructed as follows [\BOGRO, \BOSTb,
\VENc]
\plus
$$\widehat K_{N,D}(z,w)=\half\sum_{\{\gamma\}}
      \Big[k(z,\gamma w)\pm k(z,\gamma\CI w)\Big],
  \tag\NUM.\num$$
where the ``+''-sign stands for Neumann, and the ``-''-sign for
Dirichlet boundary-con\-di\-tions. $\sum_{\{\gamma\}}$ denotes the
summation over distinct conjugacy classes. Let $L$ the Selberg-operator
with eigenvalue $\Lambda (\lambda)$ on $\CF$ (where $k(z,w)$ is the
corresponding integral kernel) together with its counterpart $\widehat
L$ on $\hatCF$, and we introduce the Maass-Laplacian
$D_m=-y^2(\partial_x^2+\partial_y^2)+imy\partial_x$ on $\CF$ in
$\CL^2(\CF,m,\chi)\equiv\CL^2(\CF)$ (the space of square integrable
automorphic forms).

To each cusp there is associated an {\it Eisenstein series}
\plus
$$e(z,s,\alpha)=\sum_{\gamma\in\Gamma_\alpha\backslash\Gamma}
        y^s(\gamma z)
  \tag\NUM.\num$$
$z\in\CH$, $\Re(s)>1$, $\alpha=1,\dots,\kappa$, with $\Gamma_\alpha$ the
stabilizer of the cusp $\alpha$. In the spectral decomposition of $D_m$
on $\CL^2(\CF)$ these Eisenstein series span the continuous spectrum.

\eject
\noindent
Let us start with Dirichlet boundary-conditions and only the odd
automorphic functions with respect to $x$ survive in the spectral
expansion of the automorphic kernel. Let $f\in\CL^2(\CF)$ an odd
eigenfunction in $x$ of $D_m$ with eigenvalue $\lambda$.  A glance on
the continuous spectrum shows that the Eisenstein series $e(z,s)$ drop
out, c.f.\ according to a theorem by Venkov [\VENc]. Then
\plus
$$\allowdisplaybreaks\align
  (\widehat L_Df)(z)
            &=\half\int_{\widehat\CF}\widehat K_D(z,z')f(z')dV(z')
  \\        &=\half(Lf)(z)-\half(Lf)(-\bar z)=\Lambda(\lambda)f(z);
  \tag\NUM.\num\endalign$$
if $f\in\CL^2(\CF)$ even, similarly for Neumann boundary-conditions
\plus
$$\allowdisplaybreaks\align
  (\widehat L_Nf)(z)
            &=\half\int_{\widehat\CF}\widehat K_N(z,z')f(z')dV(z')
  \\        &=\half(Lf)(z)+\half(Lf)(-\bar z)=\Lambda(\lambda)f(z).
  \tag\NUM.\num\endalign$$
In the case of Dirichlet boundary-conditions we are thus left with the
spectral expansion of the automorphic kernel into {\it odd discrete}
eigenfunctions $\Psi_n$ on $\CH$
\plus
$$\widehat K_D(z,w)=\sum_n h(p_n)\Psi_n(z)\Psi_n(w),
  \tag\NUM.\num$$
where $\Lambda(\lambda_n)=\Lambda(p_n^2+\viert)\equiv h(p_n)$.
In the case of Neumann boundary-conditions we get the
spectral expansion of the automorphic kernel into {\it even discrete and
continuous} eigenfunctions $\Phi_n$ and $e(z,s)$, respectively, on $\CH$
\plus
$$\widehat K_N(z,w)=\sum_n h(p_n)\Phi_n(z)\Phi_n(w)
  +{1\over4\pi}\int_{-\infty}^\infty h(p)
  e(z,\bhalf+ip)\overline{e(w,\bhalf+ip)}.
  \tag\NUM.\num$$

\medskip\noindent
{\it 2.Conjugacy classes on bordered Riemann surfaces and the trace
formula.}
As usual the quantity $N_\gamma$ is called {\it norm} of an hyperbolic
$\gamma\in\Gamma$ and $N_{\gamma_0}$ will denote the norm of a
primitive hyperbolic $\gamma\in\Gamma$, and $l_\gamma=\ln N_{\gamma}$
denotes the {\it length} corresponding to a $\gamma\in\Gamma$.
$\gamma_0\in\Gamma$ is called a primitive element, if it is not a power
of any other element of $\Gamma$. Each element $\gamma\in\Gamma/
\{\pm\bbbone\}$ is thus uniquely described as $\gamma=k^{-1}\gamma_0k$
for some primitive $\gamma_0$, $n\in\bbbn$ and $k\in\Gamma/
\Gamma_{\gamma_0}$. Conjugacy classes are defined by  $\{\gamma\}
:=\{\gamma_i\in\Gamma| \gamma_i=k^{-1}\gamma k,k\in\Gamma\}$.
$\sum_{\{\gamma\}}$ denotes the summation over primitive conjugacy
classes of a particular conjugacy class $\{\gamma\}$ within the
Fuchsian group $\hatGamma$. In the following I will omit the index
``$0$'' in $\gamma_0$ if it is obvious that indeed the primitive
$\gamma_0$ is meant and no confusion can arise.

Let $\kappa$ be the number of inequivalent cups on the fundamental
domain $\hatCF$ (the number of primitive parabolic conjugacy classes in
$\widehat \Gamma$), and by $\Gamma_\rho$ the restriction of $\hatGamma$
to this inversion counterpart, where I have abbreviated
$\rho:=\gamma\CI$. In order to investigate the various conjugacy
classes for the formulation of the Selberg trace-formula for bordered
Riemann surfaces, we have to distinguish the original conjugacy classes
which appear already for closed Riemann surfaces and additional
conjugacy classes due to $\gamma\CI$. The new conjugacy classes can be
characterized by their traces. We consider first compact Riemann
surfaces, i.e.\ compact polygons as fundamental domains. The case of
the closed Riemann surfaces gives us hyperbolic and elliptic conjugacy
classes which correspond to $|\tr(\gamma)|>2$, respectively
$|\tr(\gamma)|<2$.

In the theory of symmetric spaces it is convenient to consider the
following isomorphic model of $\CH$. One defines the positive definite
symmetric matrices
\plus
$$z(x;y)=\pmatrix y+x/y^2 &x/y  \\
                  x/y     &1/y  \endpmatrix,
  \qquad(x\in\bbbr,y>0).
  \tag\NUM.\num$$
If $g\in\SL(2,\bbbr)$, then the group action has the form
\plus
$$gz(x;y)=g[z(x;y)]g^t,
  \tag\NUM.\num$$
where $g^t$ denotes the transpose of $g$. In this model it is easy to
implement the involution $\CI$ in terms of the matrix
\plus
$$\CI=\pmatrix 1  &0  \\
               0  &-1 \endpmatrix,
  \tag\NUM.\num$$
i.e.\ $\CI$ is an element in $\GL(2,\bbbr)/\{\pm\bbbone\}$.

Within this model we find, first, for $\tr(\rho)\not=0$ that
the relative centralizer is of the form
\plus
$$\pmatrix b &0 \\ 0&-b^{-1}\endpmatrix,\qquad  (\mod\pm1).
  \tag\NUM.\num$$
[Centralizers $\Gamma_\gamma$ are defined by $\Gamma_\gamma:=
\{\gamma_i\in\Gamma| \gamma_i^{-1}\gamma\gamma_i=\gamma\}$].
Therefore $\Gamma\CI\subset\hatGamma$ consists of hyperbolic
elements and the identity and since $\hatGamma$ is discrete
of a single hyperbolic element. The second case gives $\tr(\rho)=0$.
Then the relative centralizer consists of elements of the form
\plus
$$\rho_1=\pmatrix c &0 \\ 0&c^{-1}\endpmatrix,\qquad
  \rho_2=\pmatrix 0 &d \\ -d^{-1} &0\endpmatrix,\qquad
  (\mod\pm1).
  \tag\NUM.\num$$
\edef\numbd{\NUM.\num}%
$\rho_2$ is an elliptic element of order two. Thus $\gamma\CI$ consists
of hyperbolic, elliptic and the identity element. However, due to the
construction $\rho_1^n\rho_2$ $(n\in\bbbz)$ we see that we can generate
infinitely many elliptic conjugacy classes which is impossible, since
$\hatGamma$ is discrete. Therefore the relative centralizer of $\gamma
\CI$ with $\tr(\gamma\CI)=0$ consists either of hyperbolic elements and
the identity or by a single elliptic generator of order two. The
explicit computation reveals that in the compact case only the former
case is possible, the latter leading to a divergency.

The conjugacy classes of $\rho\in\hatGamma\CI$ can therefore be
distinguished in two ways [\BLCL, \BOGRO, \BOSTb] according to their
squares $\rho^2\in\hatGamma$. Let $\rho\in\hatGamma$ be primitive,
that is not a positive power of any other element of $\hatGamma\CI$.
Then
\edef\folioc{\folio}
\medskip
\item{i)} $\rho=\rho_i$, $\rho_i^2\in\{C_i\}_{\hatGamma}$, $i=1,
          \dots,n$. The $\{C_i\}_{\hatGamma}$ are the conjugacy
          classes of the $C_i$ in $\hatGamma$ which correspond to
          the closed geodesics $c_i$ on $\widehat\Sigma$.
\item{ii)} $\rho=\rho_p$, $\rho_p^2$ being a primitive element in
           $\hatGamma$ and $\rho_p^2\not=\{C_i\}
           _{\hatGamma}$.

\medskip\noindent
In the notation of Venkov [\VENc] the relative hyperbolic conjugacy
classes with $\{\rho\}$ with $\tr(\rho)=0$ correspond to the case i),
and  the relative hyperbolic conjugacy classes with $\{\rho\}$ with
$\tr(\rho)\not=0$ correspond to the case ii).

\eject
\noindent
Thus it follows that the sum over conjugacy classes for $\rho\in
\hatGamma\CI$ is divided into first the conjugacy classes of the $C_i$
in $\hatGamma$, which correspond to the closed geodesics $c_i$ on
$\widetilde\Sigma$, and second into conjugacy classes such that for all
$\rho\in\hatGamma\CI$ there is a unique description $\gamma=k^{-1}
\rho^{2n-1}k$ $(n\in\bbbn)$, for $\rho\in\hatGamma\CI$ inconjugate and
primitive, and $k\in\Gamma_{\rho^2}\backslash\widehat \Gamma$.

Let us continue by considering a non-compact polygon. The additional
conjugacy classes are again classified according to their trace. The
conjugacy class $\tr(\gamma)=2$ with corresponding non-compact Fuchsian
group $\hatGamma$ gives the already known parabolic conjugacy classes.
Each conjugacy classes with $\tr(\rho)=0$ give rise to an elliptic
transformation whose centralizer is generated by a single generator of
order two (see above). In other words, for each
$\gamma\in\hatGamma$ there exists an element $g\in\PSL(2,\bbbr)$
having the properties
\plus
$$g\rho g^{-1}=\CI,\qquad
  g\hatGamma_\rho g^{-1}
  \left\{\bbbone_2,\pmatrix 0  &a\\-1/a &0\endpmatrix
                 (\mod\pm1)\right\},
  \tag\NUM.\num$$
where $a\geq1$. These classes play the r\^ole of the parabolic classes
in the classical Selberg trace formula. Evaluating all contributions,
we can write down

\medskip\noindent
{\bf Theorem 2.1} [\BOGRO, \BOSTb, \VENc]:
\it
The Selberg trace formula on arbitrary bordered Riemann surfaces for
automorphic forms of weight $m$, $m\in\bbbz$, is given by
\plus
$$\allowdisplaybreaks\align
   \sum_{n=1}^\infty h(p_n)
  &=-{\hatCA\over16\pi^2}
   \int_0^\infty{\cosh{um\over2}\over\sinh{u\over2}}g'(u)du
   +\viert\sum_{\{\gamma\}}\sum_{k=1}^\infty
   {\chi_{\gamma}^{mk}l_\gamma g(kl_\gamma)\over\sinh{kl_\gamma\over2}}
  \\   &\quad+
   {i\over4}\sum_{\{R\}}\sum_{k=1}^{\nu -1}\chi_{R}^{mk}
   {e^{i(m-1)k\pi/\nu}\over\nu\sin(k\pi/\nu)}
   \int_{-\infty}^\infty du\,g(u)
   {e^{(m-1)u/2}(e^u-e^{2ik\pi/\nu})\over\cosh u+\cos[\pi-2(k\pi/\nu)]}
  \\   &\quad-
   \viert\sum_{\{\rho^2\}}\sum_{k=0}^\infty
   {\chi^{m(2k+1)}_\rho \chi_\CI^m l_{\rho^2}g[(k+\half)l_{\rho^2}]
    \over\cosh{\big[\half(k+\half)l_{\rho^2}\big]}}
   -\half\sum_{i=1}^n\sum_{k=1}^\infty
   {\chi_{C_i}^{mk}l_{C_i}g(kl_{C_i})\over\cosh{kl_{C_i}\over2}}
  \\   &\quad+
   {g(0)\over2}
   \left[\viert\sum_{\scriptstyle \{\rho\};\,\hatGamma_{\rho,ell}
                \atop\scriptstyle \tr(\rho)=0}\chi_\rho^{m}
   \ln\bigg({a(\rho)\over\nu(\rho)}\bigg)-\tkappa\ln2-{L\over2}\right]
   +{\tkappa\over8}h(0)
  \\   &\quad-
   {\tkappa\over4\pi}\int_{-\infty}^\infty h(p)\Psi(\bhalf+ip)dp
  +{\tkappa\over4}\int_0^\infty{g(u)\over\sinh{u\over2}}
   \bigg(1-\cosh{um\over2}\bigg)du.
    \tag\NUM.\num\endalign$$
\edef\numca{\NUM.\num}%
with the abbreviation $L=\sum_{i=1}^n  l_{C_i}$ and where the
$\lambda_n=\viert+p_n^2$ on the left run through the set of all
eigenvalues of the Dirichlet problem, and the summation on the right is
taken over all primitive conjugacy classes $R\in\hatGamma$ with
$\tr(R)<2$, $\gamma\in\hatGamma$ with $\tr(\gamma)>2$, and
$\gamma\CI\in\hatGamma$, $\tr(\rho)\not=0$. The lengths
$l_{C_i}$ are twofold degenerate, since $C_i$ and $C_i^{-1}$ both have
to be included into the sum.
\newline
$h(p)$ denotes an even function in $p$
and must has the following properties
\medskip
\item{i)} $h(p)$ is holomorphic in the strip
          $|\Im(p)|\leq\half+\epsilon$, $\epsilon>0$.
\item{ii)} $h(p)$ has to decrease faster than $|p|^{-2}$ for
           $p\to\pm\infty$.
\item{iii)} $g(u)=\pi^{-1}\int_0^\infty h(p)\cos(\pi p)dp$.
\par\nobreak\noindent
\centerline{\hfill\vrule height0.02cm depth0.3cm width0.3cm}

\rm
\eject
\baselineskip=12pt
\bigskip\goodbreak\noindent
Note that for Neumann boundary-conditions the inverse-hyperbolic
terms change their signs. In this case, however, the parabolic terms
are quite different, due to the additional presence of the continuous
spectrum represented by Eisenstein-series, see e.g.\ Ref.[\HEJb].

\noindent
$a(\rho)$ and $\mu(\rho)$ are quantities specific to the conjugacy
class of the elliptic $\gamma\in \hatGamma$ with $\tr(\rho)=0$ which
will be explained later on [c.f.\ Theorem 4.2], and we require the
following property of the multiplier system
\plus
$$\tkappa:=\sum_{\{S\}}\chi_S^m
  =\sum_{     \scriptstyle     \{\rho\};\,\hatGamma_{\rho,ell}
         \atop\scriptstyle    \tr(\rho)=0}\chi_\rho^m.
  \tag\NUM.\num$$


\bigskip\goodbreak\noindent
\glno=0                
\advance\chapno by 1   
\line{\bf III.\ Super Riemann Surfaces, Bordered Super Riemann Surfaces
      and the\hfill}
\line{\bf\phantom{\bf III.\ }%
      Selberg Supertrace Formula on Closed Super Riemann Surfaces\hfill}
\par\medskip\nobreak\noindent
{\it 1.\ Super Riemann surfaces and construction of bordered super
Riemann surfaces.} We sketch some important facts about super Riemann
surfaces. For more details I refer to Batchelor et al.\ [\BATCH, \BABR],
DeWitt [\DEW], Moore, Nelson and Polchinski [\MNP], Ninnemann [\NINN],
Rabin and Crane [\RC], and Rogers [\ROG]. Let us start with a $(1|1)$
(complex)-dimensional (not necessarily) flat superspace, parameterized
by even coordinates $Z\in\bbbc_c$ and odd (Grassmann) coordinates
$\theta\in\bbbc_a$, respectively. Let $\Li$ be the infinite dimensional
vector space generated by elements $\zeta_a$ $(a=1,2,\dots)$ with basis
$1,\zeta_a,\zeta_a\zeta_b,\dots$ $(a<b)$ and the anticommuting relation
$\zeta_a\zeta_b=-\zeta_b\zeta_a$, $\forall_{a,b}$. Every $Z\in\Li$ can
be decomposed as $Z=Z_B+Z_S$ with $Z_B\in\bbbc_c\equiv\bbbc$,
$Z_S=\sum_n{1\over n!}c_{a_1,\dots,a_n}\zeta^{a_n}\dots\zeta^{a_1}$,
with the $c_{a_1,\dots,a_n}\in\bbbc_a$ totally antisymmetric. $Z_B$ and
$Z_S$, respectively, are called the {\it body} (sometimes denoted by
$Z_B=Z_{red}$) and {\it soul} of the supernumber $Z$, respectively.
The notion of superspace and supermanifolds as introduced by
Batchelor and Bryant [\BATCH, \BABR], DeWitt [\DEW], Rabin and Crane
[\RC], and Rogers [\ROG] enables one to represent supersymmetry
transformations as pure geometric transformations in the coordinates
$Z=(z,\theta)\in\bbbc_c\times\bbbc_a$. As is well-known, a usual complex
manifold of complex dimension equal to one is already a Riemann surface.
The definition of a super Riemann surface, however, requires the
introduction of a super-conformal structure. Let us consider the
operator $D=\theta\partial_z+\partial_\theta$ (note $D^2=\partial_z$).
Further we consider a general superanalytic coordinate transformation
$\widetilde z=\widetilde z(z,\theta)$, $\widetilde\theta
=\widetilde\theta(z,\theta)$. A superanalytic coordinate
transformation is called superconformal, iff the $(0|1)$-dimensional
subspace of the tangential space generated by the action of $D$ is
invariant under such a coordinate transformation, i.e.\
$D=(D\widetilde\theta)\widetilde D$. This means that a coordinate
transformation is super-conformal iff $Dz'=\theta'D\theta'$.

To study supersymmetric field theories one needs even and odd
superfields. Here now the definition of DeWitt [\DEW] of super Riemann
manifolds conveniently comes into play. The infinity dimensional
algebra $\Li$ supplies all the required quantities. Domains in
$\bbbc^{(1|1)}$ with coordinates $(z,\theta)$ are constructed in such a
way that the entire Grassmann algebra are attached to the usual complex
coordinates. If one considers the universal family of DeWitt super
Riemann manifolds with genus $g$, then only $2g-2$ parameters of
$\Li$ are required, the remaining ones are redundant.

An important property we need in our investigations is, when a
supermanifold is split. This means that for a coordinate transformation
$Z\to Z'$ ($Z,Z'\in\Li$) the coefficient functions do not mix which
each other. Let $x$ be usual local coordinates, and $\zeta\in\Li$ local
Grassmann coordinates, then if a supermanifold is split then there is a
global isomorphism such that the coefficient functions $y$ and $\eta$
of a super-functions $F(x,\zeta)$ transform according to
\plus
$$\left.
  \aligned
  y   &=a_0(x)+a_{ij}(x){\zeta}^i{\zeta}^j+\dots,
  \\
\eta&=b_{1,i}(x){\zeta}^i+b_{3,ijk}(x){\zeta}^i{\zeta}^j{\zeta}^k+\dots,
  \endaligned\quad\to\quad\aligned
  &a_0'(x')+a_{ij}'(x'){\zeta'}^i{\zeta'}^j+\dots,
  \\
  &b_{1,i}'(x'){\zeta'}^i+b_{3,ijk}'(x')
             {\zeta'}^i{\zeta'}^j{\zeta'}^k+\dots,
  \endaligned\qquad\right\}
  \tag\NUM.\num$$
for $Z\to Z'$.
Due to a theorem of Batchelor [\BATCH] every differentiable
supermanifold is split, and in particular every complex supermanifold
of dimension $(d|1)$. The super Riemann surfaces in question can be
seen as complex a $(1|1)$-dimensional supermanifold, respectively a real
$(2|2)$-dimensional manifold,  where the coordinate transformations
are super-conformal mappings [\RC].

To generalize the uniformization theorem for Riemann surfaces to super
Riemann surfaces $\CM$, one shows that unique generalizations
$\widehatC^{(1|1)}$, $\bbbc^{(1|1)}$ and $\CH^{(1|1)}:=
\{(z,\theta)\in\bbbc^{(1|1)} |\Im(z)>0\}$ of simple connected Riemann
surfaces exist, and endows $U=\widehatC^{(1|1)}$, $\bbbc^{(1|1)}$ and
$\CH^{(1|1)}$, respectively, with a super-conformal structure, such
that the local coordinate transformations are super-conformal mappings
[\RC].

In the case of non-euclidean harmonic analysis in the context of super
Riemann surfaces we consider the group $\OSp(2,\bbbc)$ of super
conformal automorphisms on super Riemann surfaces as a natural
generalization of M\"obius transformations. They have the form
\plus
$$\multline
 \OSp(2,1;\bbbc_c^2\times\bbbc_a):=\left\{
  \gamma=\pmatrix
   a       &b     &\chi_\gamma(b\alpha-a\beta)   \\
   c       &d     &\chi_\gamma(d\alpha-c\beta)   \\
   \alpha  &\beta &\chi_\gamma(1-\alpha\beta)    \endpmatrix\right|
  a,b,c,d\in\bbbc_c;
  \hfill\\  \hfill
  \alpha,\beta\in\bbbc_a;\,ad-bc=1+\alpha\beta;
  \,\sdet \gamma=\chi_\gamma\in\{\pm1\}\Bigg\}
  \endmultline
  \tag\NUM.\num$$
($\alpha,\,\beta$ real, with the complex conjugate rules $\overline{f+g}
=\bar f+\bar g$, and $\overline{f\cdot g}=\bar f\cdot\bar g$). Its
generators are the operators $L_0$, $L_1$, $L_{-1}$, $G_{1/2}$ and
$G_{-1/2}$ of the Neveu-Schwarz sector of the super Virasoro algebra of
the fermionic string. Elements $\gamma\in \OSp(2,1;\bbbc_c^2\times\bbbc_
a)$ act on elements $x=(z_1,z_2,\xi)\in \bbbc_c^2\times\bbbc_a\setminus
\{0\}$ by matrix multiplication, i.e.\  $x'=\gamma x$.
\edef\folioa{\folio}%
By means of a local coordinate system $(z,\theta)=(z_1/z_2,\xi/z_2)$
and the requirements of superconformal transformation the local
coordinate transformations are fixed and the super M\"obius
transformations explicitly have the form [\BMFSb, \GROd, \NINN, \RC,
\UEYA]
\plus
$$z'={az+b\over cz+d}+\theta{\alpha z+\beta\over(cz+d)^2},\qquad
  \theta'={\alpha+\beta z\over cz+d}
  +{\chi_\gamma\theta\over cz+d}.
  \tag\NUM.\num$$
The $\chi_\gamma$ with $\chi_\gamma=\pm1$ lead to the description of
spin structures on a super Riemann surface. The transformation factor
of the $D$ operator yields to
\plus
$$F_\gamma:=(D\theta')^{-1}=\chi_\gamma(cz+d+\delta\theta),
  \tag\NUM.\num$$
with $\delta=\chi_\gamma\sqrt{1+\alpha\beta}\,(\alpha d+\beta c)$. This
general super-M\"obius transformation does mix the coefficient functions
 of superfunctions $F\in\Li$. Since we required that the super Riemann
surfaces in question is split, the odd quantities $\alpha,\beta$ are
not necessary and can be omitted. It is sufficient to consider
transformations $\gamma\in\OSp(2,1)$ with $\alpha=\beta=0$ and the
characters $\chi_\gamma$ which describe spin structures. Furthermore
$\gamma$ and $-\gamma$ describe the same transformation. We thus have
that the automorphisms on $\SCH$ are given by
\plus
$$\Aut\SCH={\OSp(2|1,\bbbr)\over\{\pm\bbbone\}}.
  \tag\NUM.\num$$
and a super Fuchsian group $\Gamma$ denotes a discrete subgroup of
$\Aut\SCH$. Therefore we obtain for the transformations $z\to z'$ and
$\theta\to\theta'$ [\BMFSb, \NINN]
\plus
$$z'={az+b\over cz+d},\qquad
  \theta'={\chi_\gamma\theta\over cz+d}\enspace,
  \tag\NUM.\num$$
[here $F_\gamma=\chi_\gamma(cz+d)$]. $M_{\xi=0}$ corresponds to the
usual Riemann surface $M_{red}$ with some spin-structure, since a
$\gamma\in\Aut\SCH$ is fixed by a $\PSL(2,\bbbr)$ transformation and a
character $\chi_\gamma=\pm1$. The properties of the odd coordinates is
determined by the properties of $M_{red}$ and $\theta$ is the cut of a
spinor-bundle.

\medskip\noindent
{\it 2.\ Dirac-Laplace operators and conjugacy classes on super Riemann
surfaces.}
We need some further ingredients. Let us introduce the quantities
$N_\gamma$ and $l_\gamma$
\plus
$$2\cosh{l_\gamma\over2}=N_\gamma^{1/2}+N_\gamma^{-{1/2}}
  =a+d+\chi_\gamma\alpha\beta.
  \tag\NUM.\num$$
$N_\gamma$ is called {\it norm} of an hyperbolic $\gamma\in\Gamma$ in a
(general) super Fuchsian group, and $N_{\gamma_0}$ will denote the norm
of a primitive hyperbolic $\gamma\in\Gamma$, and $l_\gamma=\ln
N_{\gamma}$ denotes the {\it length} corresponding to a
$\gamma\in\Gamma$ and all notions from the bosonic case are interpreted
in a straightforward way into their super generalization. Each element
$\gamma\in\Gamma/ \{\pm\bbbone\}$ is thus uniquely described as
$\gamma=k^{-1} \gamma_0k$ for some primitive $\gamma_0$, $n\in\bbbn$
and $k\in\Gamma/ \Gamma_{\gamma_0}$. For $\OSp(2,\bbbr)/\{\pm\bbbone\}$
in homogeneous coordinates a hyperbolic transformation is always
conjugate to the transformation $z'=N_\gamma z,\ \theta'=
\chi_\gamma\sqrt{N_\gamma} \,\theta$, or in matrix representation
\plus
$$\hbox{hyperbolic $\gamma\in\Gamma$ conjugate to}\qquad
  \pmatrix  N_\gamma^{1/2}  &0                 &0 \\
            0               &N_\gamma^{-{1/2}} &0 \\
            0               &0                 &\chi_\gamma
  \endpmatrix.
  \tag\NUM.\num$$
Hyperbolic transformations are also called dilatations.

The generators of a particular super Fuchsian group of a super
Riemann surface with genus $g$ obey the constraint
\plus
$$(\gamma_0\gamma_1^{-1}\dots\gamma_{2g-2}\gamma_{2g-1}^{-1})
  (\gamma_0^{-1}\gamma_1\dots\gamma_{2g-2}^{-1}\gamma_{2g-1})=
  \bbbone_{2|1}.
  \tag\NUM.\num$$
\edef\numbe{\NUM.\num}%
In order to construct explicitly a metric on $\SCH$ one starts with the
super Vierbeins in flat superspace and performs a super Weyl
transformation [\HOWE] to obtain the metric $ds^2=dq^a{_a}g_bdq^b$ in
$\SCH$ [\UEYA]. The scalar product has the form
\plus
$$(\Phi_1,\Phi_2)=\int_\SCH{dzd\bar zd\theta d\bar\theta\over 2Y}
   \Phi_1(Z)\bar\Phi_2(Z)
   \equiv\int_\SCH dV(Z)\Phi_1(Z)\bar\Phi_2(Z),
  \tag\NUM.\num$$
for super functions $\Phi_1,\Phi_2\in L^2(\SCH)$ and $Y=y+
i\theta\bar\theta/2=y+\theta_1\theta_2$ $(\theta=\theta_1+i\theta_2$).
We have one even and one odd point pair invariant given by
[\BMFSb, \MUYb, \UEYA]
$$\allowdisplaybreaks\align
  R(Z,W)&={|z-w-\theta\nu|^2\over YV}
  \tag\NUM.\num\\    \global\plus
  r(Z,W)&=i{2\theta\bar\theta+(\nu+\bar\nu)(\theta-\bar\theta)\over4Y}
          +i{2\nu\bar\nu+(\theta+\bar\theta)(\nu-\bar\nu)\over4V}
  \\  &\qquad\qquad\qquad
     +{(\nu-\bar\nu)(\theta-\bar\theta)\Re(z-w-\theta\nu)\over4YV}
  \tag\NUM.\num a\\  &
  ={(\theta_1-\nu_1)\theta_2\over y}+{(\nu_1-\theta_1)\nu_2\over v}
     +{\theta_2\nu_2\Re(z-w-\theta\nu)\over4YV}
  \tag\NUM.\num b\endalign$$
($Z,W\in\SCH$, $W=(w,\nu)=(u+iv,\nu_1+i\nu_2), V=v+i\nu\bar\nu/2$) as
derived form classical mechanics on the Poincar\'e super upper
half-plane [\AOK, \MUYb, \UEYA].
We introduce the Dirac-Laplace operators $\square_m$
and $\hsquare_m$, respectively [\AOK, \BMFSb]
\plus
$$\square_m=2YD\bar D+im(\bar\theta-\theta)\bar D,\qquad
  \hsquare_m=2YD\bar D+{im\over2}(\bar\theta-\theta)(D+\bar D),
  \tag\NUM.\num$$
and $\square_m$ and $\hsquare_m$ are related by a linear isomorphism
$\square_m=Y^{-m/2}(\hsquare_m+im/2)Y^{m/2}$. Particularly we have for
$m=0$
\plus
$$\hsquare_0=\square_0\equiv\square
  =2Y(\partial_\theta\partial_{\bar\theta}
    +\theta\bar\theta\partial_z\partial_{\bar z}
    +\theta\partial_{\bar\theta}\partial_z
    -\bar\theta\partial_\theta\partial_{\bar z}).
  \tag\NUM.\num$$
With the notation $-\Delta_m=-4y^2\partial_z\partial_{\bar z}
+imy\partial_x=-y^2(\partial_x^2+\partial_y^2)+imy\partial_x$
we obtain for a super function
\plus
$$\Psi(Z,\bar Z)=A(z,\bar z)+{\theta\bar\theta\over y}B(z,\bar z)
                  +{1\over\sqrt{y}}\Big(\theta\chi(z,\bar z)
                  +\bar\theta\widetilde\chi(z,\bar z)\Big)
  \tag\NUM.\num$$
the following equivalence [\BMFSb, \GROd, \NINN]
\plus
$$\allowdisplaybreaks\align
  \hsquare_m\Psi(Z,\bar Z)
  &=s\Psi(Z,\bar Z)
  \\
  &\Longleftrightarrow\left\{\aligned
  &-\Delta_{m}A(z,\bar z)=s(s+i)A(z,\bar z),
  \\
  &B(z,\bar z)={s\over2}A(z,\bar z),
  \\
  &\bigg(s-{im\over2}\bigg)\widetilde\chi(z,\bar z)
  =-2y\partial_{\bar z}\chi(z,\bar z)+{i\over2}(m+1)\chi(z,\bar z)
  \\
  &-\Delta_{(m+1)}\chi(z,\bar z)=\bigg(\viert+s^2\bigg)\chi(z,\bar z).
  \endaligned\right.
  \tag\NUM.\num\endalign$$
\edef\numba{\NUM.\num}%
An explicit solution of Eq.(\numba) for $m=0$ on the entire $\SCH$
is given by [\MUYb]
$$\allowdisplaybreaks\align
  \Phi_{p,k}(z,\bar z,\theta,\bar\theta)
  &=\sqrt{2i\sinh\pi p\over \pi^3}\,
  \bigg(1-i{1+2ip\over4y}\theta\bar\theta\bigg)
  \sqrt{y}\,e^{ikx}K_{ip}(|k|y)
  \tag\NUM.\num\\    \global\plus
  \phi_{p,k}(z,\bar z,\theta,\bar\theta)
  &=\sqrt{\cos[\pi(c+ip)]\over2\pi^2(c+ip)^{\sigma_k-1}}
    {e^{ikx}\over\sqrt{y}}
  \\   &\qquad\times
  \Big[\theta W_{\sigma_k/2,c+ip}(2|k|y)
    +i(c+ip)^{\sigma_k}\bar\theta W_{-\sigma_k/2,c+ip}(2|k|y)\Big]
  \tag\NUM.\num\endalign$$
with $s=-i(\half+ip)$, $\sigma_k=\sign(k)$, $(k\not=0)$, and
$c\in\bbbr$, $|c|\leq\half$. $K_\nu$ and $W_{\mu,\nu}$ denote modified
Bessel- and Whittaker-functions, respectively. Due to the particular
form of the differential equation for $\Phi(Z,\bar Z)$ we see that the
solutions can be characterised by their parity with respect to the
coordinate $x$, i.e.\ they can have even and odd parity with respect to
$x$.

I have proposed in II similarly as for the hyperbolic $T\in\Gamma$,
elliptic and parabolic $T\in\Gamma$, and appropriate super fundamental
domains $\SCF$, a decomposition of an appropriate $T\in\Gamma$ as
follows [\GROe]
\plus
$$\multline
  (\hbox{$T\in\Gamma$ conjugate to)}\qquad
  \gamma\times R\times S
  =\pmatrix N_\gamma^{1/2}  &0                 &0           \\
            0               &N_\gamma^{-{1/2}} &0           \\
            0               &0                 &\chi_\gamma
  \endpmatrix
  \\  \times      \pmatrix
  \cos\phi &-\sin\phi &0       \\
  \sin\phi &\cos\phi  &0       \\
  0        &0         &\chi_R
  \endpmatrix\cdot\pmatrix
  1 &n &0        \\
  0 &1 &0        \\
  0 &0 &\chi_S   \endpmatrix,
  \endmultline
  \tag\NUM.\num$$
with $n\in\bbbn$ and $0<\phi<\pi$, and $\gamma,\,R$ and $S$,
respectively, denote hyperbolic, elliptic and parabolic transformations,
acting by matrix multiplication [c.f.\ p.\folioa]. The body $\CF$ of a
fundamental domain $\SCF$ has according to [\HEJb] $4g+2s+2\kappa$
sides, the boundaries being geodesics, of course. We also maintain the
notion of $\chi_T$ irrespective, whether $T\in\Gamma$ is hyperbolic,
elliptic or parabolic, respectively, and we choose $\chi_T$ according
to the spin structure of the super Riemann surface in question. For a
super Riemann surface of genus $g$ there are obviously $2^{(\#
generators)}=2^{(2g+s+\kappa)}$ possible spin structures.

The constraint (\numbe) is altered due to the presence of of elliptic
fixed points and cusps according to [\HEJb, \VENc]
\plus
$$(\gamma_0\gamma_1^{-1}\dots\gamma_{2g-2}\gamma_{2g-1}^{-1})
  (\gamma_0^{-1}\gamma_1\dots\gamma_{2g-2}^{-1}\gamma_{2g-1})
  R_1\dots R_s S_1\dots S_\kappa=\bbbone_{2|1}.
  \tag\NUM.\num$$

\medskip\noindent
{\it 3.\ Construction of bordered super Riemann surfaces.}
Because it is sufficient to consider super Riemann surfaces without
odd parameters we can propose a construction of a bordered super
Riemann surface. To construct a bordered super Riemann surface we take
the construction of a usual bordered Riemann surface and endow it with
the Grassmann algebra $\Li$. Because we know how to define a
closed super Riemann surface, we take $\widehat\Sigma$ and enlarge it
to $\widehat\Sigma^{(1|1)}$ together with its corresponding super
Fuchsian group $\hatGamma^{(1|1)}$ constructed from $\hatGamma$ and the
super fundamental domain $\hatSCF$. A convenient way to introduce the
super-analogue of the involution $\CI$ turns out to be the {\it super
involution}
\plus
$$\left.\aligned
  \CI Z      &=\CI(z,\theta)=(-\bar z,-i\bar\theta),
  \\
  \CI\bar Z  &=\CI(\bar z,\bar\theta)=(-z,-i\theta)
  \endaligned\qquad\qquad\right\}
  \tag\NUM.\num$$
\edef\numbb{\NUM.\num}%
respectively $\CI(z,\theta_1,\theta_2)=(-\bar z,-i\theta_1,i\theta_2)$.
It has the properties
\plus
$$\CI D=i\bar D,\qquad   \CI\bar D=iD.
  \tag\NUM.\num$$
Note $\CI^4Z=Z$ and $\CI^4D=D$. Furthermore for the Dirac-Laplace
operator $\hsquare_m$ we have
\plus
$$\CI\,\hsquare_m=\hsquare_{-m}=\overline{\hsquare_m}.
  \tag\NUM.\num$$

Similarly as for the usual bordered Riemann surface where $\Sigma=
\widehat\Sigma\backslash\CI$, we then define the bordered super Riemann
surface $\Sigma^{(1|1)}$ as $\Sigma^{(1|1)}=\widehat\Sigma^{(1|1)}
\backslash\CI$. The corresponding discs  $d_1^{(1|1)}, \dots,$
$d_n^{(1|1)}$ then are super-conformal non-overlap\-ping superdiscs
seen as usual conformal non-overlap\-ping discs endowed with the
Grassmann algebra $\Li$. The particular form of the involution (\numbb)
enables us to work directly on the fundamental domains $\hatSCF$. The
super Fuchsian group $\hatGamma$ is consequently a symmetric super
Fuchsian group.

\medskip\noindent
{\it 4.\ The Selberg supertrace formula for hyperbolic conjugacy
classes.}
Turning to the Selberg supertrace formula, let us introduce the
Selberg super operator $L$ by [\BMFSa-\BASCH, \GROd]
\plus
$$\left.\aligned
  (L\phi)(Z)&=\int_{\SCH}dV(W)k_m(Z,W)\phi(W),
  \\
  k_m(Z,W)&=J^m(Z,W)\big\{\Phi[R(Z,W)]+r(Z,W)\Psi[R(Z,W)]\big\},
  \\
  J^m(Z,W)&=\left({z-\bar w-\theta\bar\nu
          \over\bar z-w-\bar\theta\nu}\right)^{m/2}.
  \endaligned\qquad\qquad\right\}
  \tag\NUM.\num$$
$k_m(Z,W)$ is the integral kernel of an operator valued function of the
Dirac-Laplace operator $\square_m$ (respectively $\hsquare_m$), and
$\Phi$ and $\Psi$ are sufficiently decreasing functions at infinity.
Note $J^m(\gamma Z,\gamma W)=j(\gamma,Z)J^m(Z,W)j^{-1}(\gamma, W)$
with $j(\gamma,Z)$ given by $j(\gamma,Z)=(F_\gamma/|F_\gamma|)^m$, where
$F_\gamma=D\theta'$ [\GROd, \NINN].
We have $j(\gamma\sigma,Z)=j(\gamma,\sigma Z)j(\sigma,Z)$ ($\forall
\gamma,\sigma\in\Gamma$ and $Z\in\SCH$).
A superautomorphic form $f(Z)$ is then defined by [\BMFSa, \GROd]
$f(\gamma Z)=j(\gamma,Z)f(Z)$ $(\forall\gamma\in\Gamma)$.
The super-automorphic kernel is defined as
\plus
$$K(Z,W)=\half\sum_{\{\gamma\}}k_m(Z,\gamma W)j(\gamma,W),
  \tag\NUM.\num$$
(``$\half$'' because both $\gamma$ and $-\gamma$ have to be included in
the sum) i.e.\ $(L\phi)(z)=[h(\square_m)](z)$. $L$ is acting on
super-automorphic functions $f(Z)$.

\eject
\baselineskip=11pt
\noindent
For the point pair invariants we find for the action of $\CI$
\plus
$$\left.\aligned
  R(Z,\CI W)&=R(\CI Z,W)
  \\
  r(Z,\CI W)&=\overline{r(\CI Z,W)},
  \endaligned\qquad\qquad\right\}
  \tag\NUM.\num$$
furthermore $J(Z,\CI W)=\overline{J(\CI Z,W)}$, and due to the
construction of $k_m$
\plus
$$k_m(Z,\CI W)=\overline{k_m(\CI Z, W)}.
  \tag\NUM.\num$$

Let $f$ be a super-automorphic function with $f(\gamma Z)=j(\gamma,Z)
f(Z)$ and $g=Lf$. Let $\SCFg$ a fundamental domain of $\gamma\in\Gamma$
whose body equals $\rSCF=\CF$ (and is constructed in the same sense as
the generalization $\SCH$ of $\CH$). The expansion into hyperbolic
conjugacy classes yields
\plus
$$\allowdisplaybreaks\align
  \str(L)
  &=\int_\SCFg dV(Z)K(Z,Z)
  \\
  &=\int_\SCFg\sum_{\gamma\in\Gamma}k_m(Z,\gamma Z)dV(Z)
  ={i^m\over2}\CA\,\Phi(0)
  +\sum_{     \scriptstyle \{\gamma\}
              \atop\scriptstyle \str(\gamma)+\chi_\gamma>2}
  \chi_\gamma^mA(\gamma).
  \tag\NUM.\num\endalign$$
Here I have assumed without loss of generality $a+d\geq0$ for a
$\gamma\in\hatGamma$, since $\Aut\SCH=\OSp(2|1,\bbbr)/\{\pm\bbbone\}$.
The first term corresponds to the identity transformation (zero-length
term) and the second $A(\gamma)$ is given by
\plus
$$A(\gamma)=\chi_\gamma^{-m}\int_\SCFg
  k_m(Z,\gamma Z)j(\gamma,W)dV(Z).
  \tag\NUM.\num$$
In Refs.[\BMFSb, \GROd] these two terms corresponding to the identity
transformation and hyperbolic conjugacy classes, respectively, were
calculated, i.e.\ I have discussed in detail

\medskip\noindent
{\bf Theorem 3.1} [\BMFSa-\BASCH, \GROd]:
\it
The Selberg supertrace formula for $m$-weighted Dirac-Laplace operators
on closed super Riemann surfaces for hyperbolic conjugacy classes is
given by:
\plus
$$\multline
  \!\!\!\!
  \sum_{n=0}^\infty\bigg[h\bigg({1+m\over2}+ip_n^{(B)}\bigg)
                        -h\bigg({1+m\over2}+ip_n^{(F)}\bigg)\bigg]
  =-{\CA(\CF)\over4\pi}\int_0^\infty{g(u)-g(-u)\over\sinh{u\over2}}
  \cosh\bigg({um\over2}\bigg)\,du
  \hfill\\
  +\sum_{\{\gamma\}}\sum_{k=1}^\infty
  {l_\gamma\chi_\gamma^{mk}\over2\sinh{kl_\gamma\over2}}
  \bigg[g(kl_\gamma)+g(-kl_\gamma)
  -\chi_\gamma^k\bigg(g(kl_\gamma)e^{-kl_\gamma/2}
           +g(-kl_\gamma)e^{kl_\gamma/2}\bigg)\bigg].
  \endmultline
  \tag\NUM.\num$$
\edef\numca{\NUM.\num}%
The test function $h$ is required to have the following properties
\medskip
\item{i)} $h({1+m\over2}+ip)\in C^\infty(\bbbr)$,
\item{ii)} $h({1+m\over2}+ip)$ need not to be an even function in $p$,
\item{iii)} $h(p)$ vanishes faster than $1/|p|$ for $p\to\pm\infty$.
\item{iv)} $h({1+m\over2}+ip)$ is holomorphic in the strip $\Im(p)\leq1
           +{m\over2}+\epsilon$, $\epsilon>0$, to guarantee absolute
           convergence in the summation over $\{\gamma\}$.
\par\nobreak
\line{\hfill\vrule height0.02cm depth0.3cm width0.3cm}

\rm
\eject
\medskip\noindent
The above Selberg supertrace formula (\numca) is valid for discrete
hyperbolic conjugacy classes and in this case the noneuclidean area of
the (``bosonic'') fundamental domain is $\CA=4\pi(g-1)$. The Fourier
transformation $g$ of $h$ is given by
\plus
$$\allowdisplaybreaks\align
  g(u)&={1\over2\pi}\int_{-\infty}^\infty
        h\bigg({1+m\over2}+ip\bigg)e^{-iup}dp
  \\    &
  =\viert\int_{4\sinh^2{u\over2}}^\infty {dx\over(x+4)^{m/2}}
  \left\{{\Psi(x)+2(e^u-1)\Phi'(x)\over\sqrt{x-4\sinh^2{u\over2}}}
   \big[\alpha_+^m(x,u)+\alpha_-^m(x,u)\big]
  \right.\\   &\left.
   \vphantom{{\Psi(x)+2(e^u-1)\Phi'(x)\over\sqrt{x-4\sinh^2{u\over2}}}}
   \qquad\qquad\qquad\qquad\qquad\qquad
  -ime^{u/2}\Phi(x){\alpha_+^m(x,u)-\alpha_-^m(x,u)\over x+4}\right\},
  \tag\NUM.\num\endalign$$
\edef\numcm{\NUM.\num}%
where $\alpha_{\pm}^m(x,u)= \left(\pm\sqrt{x-4\sinh^2{u\over2}}
-2i\cosh{u\over2}\right)^{m/2}$. Specific trace formul\ae, in
particular for the heat kernel were considered by Aoki [\AOK], Oshima
[\OSH], Yasui [\MUYa, \MUYb] and Uehara and Yasui [\UEYA], as well as an
explicit evaluation for the energy dependent resolvent kernel for the
operator $\hsquare^2$ [\AOK, \OSH]. From Eq.(\numcm) an explicit formula
for $\Phi(x)$ can be derived [\GROd] which has the form
\plus
$$i^m\Phi(x)={1\over\pi\sqrt{x+4}}\int_x^\infty
  {dy\over\sqrt{y+4}}\int_{-\infty}^\infty Q_1'(y+t^2)
  \left({\sqrt{y+t^2+4}-t\over\sqrt{y+t^2+4}+t}\right)^{m/2}dt,
  \tag\NUM.\num$$
\edef\numcd{\NUM.\num}%
with $Q_1(w)=2\coth{u\over2}[g(u)-g(-u)]$, $w=4\sinh^2{u\over2}$.
Let us consider the combination
\plus
$$\multline
  g(u)e^{-u/2}-g(-u)e^{u/2}={i^m\over2}\sinh{u\over2}
  \\     \times
  \int_{-\infty}^\infty d\xi
  \left({\sqrt{w+4}+i\xi\over\sqrt{w+4}-i\xi}\right)^{m/2}
  \bigg[4\Phi'(w+\xi^2)-\Psi(w+\xi^2)\bigg].
  \endmultline
  \tag\NUM.\num$$
We define $Q_3(w)=2[g(u)e^{-u/2}-g(-u)e^{u/2}]/\sinh{u\over2}$
and obtain the general inversion formula for $\Psi(x)$
\plus
$$i^m\Psi(x)=4i^m\Phi'(x)+{1\over\pi}
  \int_{-\infty}^\infty Q_3'(x+t^2)
  \left({\sqrt{x+4+t^2}-t\over\sqrt{x+4+t^2}-t}\right)^{m/2}dt.
  \tag\NUM.\num$$
\edef\numce{\NUM.\num}%
Alternatively, this can be rewritten as
\plus
$$i^m\Psi(x)=-{i^m\Phi(x)\over2(x+4)}
  +{1\over\pi}\int_{-\infty}^\infty
  \left({\sqrt{x+4+t^2}-t\over\sqrt{x+4+t^2}-t}\right)^{m/2}
  \bigg[Q_3'(x+t^2)-{Q_1'(x+t^2)\over x+4}\bigg]dt.
  \tag\NUM.\num$$
For $m=0$ we obtain simple inversion formul\ae\ for
$\Phi(t)$ and $\Psi(t)$, respectively
\plus
$$\Phi(t)=-{1\over\pi}\int_t^\infty{Q_1(w)dw\over(w+4)\sqrt{w-t}},
  \qquad
  \Psi(t)=-{1\over2\pi}\int_t^\infty{Q_2'(w)dw\over\sqrt{w-t}},
  \tag\NUM.\num$$
with $Q_2(w)=2[g(u)e^{-u/2}+g(-u)e^{u/2}]/\cosh{u\over2}$.
The incorporation of the elliptic and parabolic conjugacy classes was
discussed in Ref.[\GROe] and is not repeated here.


\eject
\baselineskip=12pt
\bigskip\goodbreak\noindent
\glno=0                
\advance\chapno by 1   
\line{\bf IV.\ The Selberg Supertrace Formula for Bordered Super
      Riemann Surfaces\hfill}
\par\medskip\nobreak\noindent
I first proceed by considering the Selberg supertrace formula where the
body of the underlying fundamental domain is compact, and second where
it is non-compact. Let us consider the super-automorphic Selberg
operator with Dirichlet boundary-conditions
\plus
$$\allowdisplaybreaks\align
  (\widehat Lf)(Z)
  &=\viert\int_{\SCH}dV(W)[k_m(Z,W)-k_m(Z,\CI W)]f(W)
  \\
  &=\viert\sum_{\{\gamma\}}
    \int_{\gamma\hatSCFg}dV(W)[k_m(Z,W)-k_m(Z,\CI W)]f(W)
  \\
  &=\half\int_{\hatSCFg}dV(W) K(Z,W)f(W),
  \tag\NUM.\num\endalign$$
where
\plus
$$K(Z,W)=\half\sum_{\{\gamma\}}[k_m(Z,\gamma W)-k_m(Z,\gamma\CI W)]
  \tag\NUM.\num$$
is the super-automorphic kernel on bordered super Riemann surfaces.
Now we have for a superfunction $\phi$ which is odd with respect to $x$
\plus
$$\allowdisplaybreaks\align
  \half&\int_{\hatSCFg}dV(W) K(Z,\CI W)\phi(W)
  \\   &=\viert\sum_{\{\gamma\}}\int_{\hatSCFg}dV(W)
  k_m(Z,\gamma \CI W)\phi(W)
  \\   &=-\viert\sum_{\{\gamma\}}\int_{\hatSCFg}dV(\CI W)
     k_m(Z,\gamma W)\phi(W)
  \\   &=-\viert\sum_{\{\gamma\}}\int_{\gamma\hatSCFg}dV(\CI W)
    k_m(Z,W)\phi(W)
  \\   &=\half\int_{\SCH}dV(W)k_m(Z,\CI W)\phi(W)
  \\   &=\half\int_{\SCH}dV(W)\overline{k(\CI Z,W)}\phi(W)
   =\half\overline{(L\bar\phi)(\CI Z)},
  \tag\NUM.\num\endalign$$
due to the properties of the super Selberg operator. Let now $\Phi$ be
an eigenfunction of $\,\hsquare_m$ which is odd with respect to $x$,
i.e.\  $\hsquare_m\Phi=s\Phi$. Then $\overline{s\Phi}=\bar
s\bar\Phi=\hsquare_{-m}\bar\Phi$ and $\bar\Phi$ is an odd eigenfunction
of $\hsquare_{-m}$ with eigenvalue $\bar s$. Denote by $\widehat L$ the
Selberg super operator on the super Riemann surface $\widehat\Sigma$;
let $(L\phi)(Z)=\Lambda(s)\phi(Z)$ and $\overline{(L\bar\phi)(\CI Z)}
=\overline{\Lambda'(\bar s)}\phi(\CI Z)$ on $\Sigma$ and $\CI\Sigma$,
respectively. Then
\plus
$$\allowdisplaybreaks\align
  (\widehat L\phi)(Z)
  &=\half(L\phi)(Z)-\half\overline{(L\bar\phi)(\CI Z)}
  \\
  &=\half\Lambda(s)\phi(Z)-\half\overline{\Lambda'(\bar s)}\phi(\CI Z)
   =\half\Big[\Lambda(s)+\Lambda'(s)\Big]\phi(Z).
  \tag\NUM.\num\endalign$$
The equivalence relation (\numba) shows that the eigenvalue problem for
the operator $\hsquare_m$ is closely related to the eigenvalue problem
of the operator $-\Delta_{m}$, both for eigenfunctions which are even
or odd  with respect to $x$. Now, an odd eigenfunction of $-\Delta_{m}$
is also an odd eigenfunction of $-\Delta_{-m}$, and the solution of the
corresponding differential equations depend only on $m^2$ but not on
$m$ [\HEJa, pp.266-68;\ \ELS, pp.203-5], hence, the spectrum depends
only on $|m|$ (compare also [\BOGRO]). Therefore we conclude that a
with-respect-to-$x$ odd eigenfunction of $\square_m$ is also a
with-respect-to-$x$ odd eigenfunction of $\CI\,\square_m$ with the
eigenvalue $\bar s$, furthermore $\Lambda=\Lambda'$ [\BOGRO], and we
can infer [together with the usual identification
$h(p)=\Lambda(\half+ip)]$
\plus
$$(\widehat L\phi)(Z)=h(p)\phi(z).
  \tag\NUM.\num$$
Let $Z_\Gamma(\gamma)$ the centralizer of a $\gamma\in\Gamma$.
For $\str(\widehat L)$ we obtain on the one hand
\plus
$$\str(\widehat L)=\sum_n\Big[h(p_n^{(B)})-h(p_n^{(F)})\Big],
  \tag\NUM.\num$$
where $s_n^{(B,F)}=\half+ip_n^{(B,F)}$ are the bosonic and fermionic
eigenvalues, respectively, of $\square_m$. [According to Eq.(\numba) we
should consequently write $s=-i(\half+ip)$, which looks, however,
somewhat artificial and is therefore not adopted.] On the other we have
\plus
$$\allowdisplaybreaks\align
  \str(\widehat L)
  &=\half\int_{\hatSCFg}dV(W) K(Z,Z)
   \\
  &=\viert\sum_{\{\gamma\}}
  \int_{\hatSCFg}[k_m(Z,\gamma Z)-k_m(Z,\gamma\CI Z)]dV(Z),
  \tag\NUM.\num\endalign$$
where $\hatSCF(\gamma)$ denotes the fundamental region for the super
Fuchsian group $Z_\Gamma(\gamma)$, the centralizer of $\gamma\in\Gamma$.

\medskip\noindent
{\it 1) $\hrSCF$ is compact.}
For convenience we set $\rho=\gamma\CI$ and use the classification of
the inverse-hyperbolic transformations according to $\rho\in\bGamma
\CI$, respectively, $\rho^2\in\bGamma$. We generalize the result of the
conjugacy classes for the usual case of bordered Riemann surfaces and
consider the two cases i) and ii) for the conjugacy classes in
$\gamma\CI$ (c.f.\ p.\folioc). The expansion into the conjugacy classes
yields for the Selberg super operator for Dirichlet boundary-conditions
[c.f.\ the discussion following Eq.(\numbd)]
\plus
$$\allowdisplaybreaks\align
  \str(\widehat L)&=
  \half\int_\hatSCFg\sum_{\{\gamma\}}
  \Big[k_m(Z,\gamma Z)-k_m(Z,\gamma\CI Z)\Big]dV(Z)
  \\   &
  ={\hatCA\over4}\,\Phi(0)
  +\half\sum_{\{\gamma\}}\int_{\hatSCF(\gamma)} k_m(Z,\gamma Z)
  -
  \half\sum_{\{\rho\};\,\hatGamma_{\rho,hyp}}
  \int_{\hatSCF(\rho)} k_m(Z,\rho Z).
  \\   &
  \tag\NUM.\num\endalign$$
Let us consider the involution term. We obtain
\plus
$$\allowdisplaybreaks\align
  \sum_{\gamma\in\bGamma}
  &\int_\hatSCFg dV(Z)k_m[Z,\CI Z)]
  \\   &=\sum_{\rho\in\bGamma\CI}\int_\hatSCFg dV(Z)k_m(Z,\rho Z)
   =:\sum_{\rho\in\bGamma\CI}A(\rho)
  \\   &=\sum_{\rho_p}\sum_{k=0}^\infty A(\rho_p^{2k+1})
   +\sum_{i=1}^n\sum_{\rho_i}\sum_{k=0}^\infty A(\rho_i^{2k+1}).
  \tag\NUM.\num\endalign$$
Now observe
\plus
$$k_m(\gamma Z,\rho^{2k+1}\gamma Z)
  =k_m(Z,\gamma^{-1}\rho^{2k+1}\gamma Z)
   j(\gamma,Z)j^{-1}(\gamma,\gamma^{-1}\rho^{2k+1}\gamma Z),
  \tag\NUM.\num$$
and
\plus
$$\allowdisplaybreaks\align
  j(\rho^{2k+1}\gamma,Z)
  &=j(\gamma,Z)j(\rho^{2k+1},\gamma Z)
   =j(\gamma\cdot\gamma^{-1}\rho^{2k+1}\gamma, Z)
  \\   &=j(\gamma,\gamma^{-1}\rho^{2k+1}\gamma Z)
   j(\gamma^{-1}\rho^{2k+1}\gamma, Z).
  \tag\NUM.\num\endalign$$
We then get
\plus
$$\allowdisplaybreaks\align
  \sum_\rho\sum_{k=0}^\infty A(\rho^{2k+1})
   &=\sum_{\{\rho\}}
    \sum_{\sigma\in\{\rho\}}\sum_{k=0}^\infty
    \int_{\hatSCF}dV(Z) k_m(Z,\sigma^{2k+1}Z)
  \\   &=\sum_{\{\rho\}}\sum_{\gamma\in\Gamma_{\rho^2}\backslash\bGamma}
    \sum_{k=0}^\infty\int_{\hatSCF}dV(Z)
    k_m(Z,\gamma^{-1}\rho^{2k+1}\gamma Z)
  \\   &=\sum_{\{\rho\}}\sum_{\gamma\in\Gamma_{\rho^2}\backslash\bGamma}
    \sum_{k=0}^\infty \int_{\hatSCF}dV(Z)
    j(\rho^{2k+1},\gamma Z)k(\gamma Z,\rho^{2k+1}\gamma Z)
  \\   &=\sum_{\{\rho\}}\sum_{k=0}^\infty
    \sum_{\gamma\in\Gamma_{\rho^2}\backslash\bGamma}
    j(\rho^{2k+1},Z)\int_{\gamma\hatSCF}dV(Z)k_m(Z,\rho^{2k+1} Z)
  \\   &=\sum_{\{\rho\}}\sum_{k=0}^\infty
    \int_{\Gamma_{\rho^2}\backslash\CH}dV(Z)
    j(\rho^{2k+1},Z)k_m(Z,\rho^{2k+1} Z).
  \tag\NUM.\num\endalign$$
By an overall conjugation in $\OSp(2|1,\bbbr)$ we can arrange for
$\gamma$ to be a dilatation, i.e.\ $\rho z=-\sqrt{N}\,\bar z$
and $\nu_1=\rho\theta_1=-\chi_\rho N^{1/4}\theta_1$,
$\nu_2=\rho\theta_2=-\chi_\rho N^{1/4}\theta_2$. Similarly as in the
usual hyperbolic case [\GROd] we find for the two-point invariants
[$M=N^{k+1/2}$]
$$\allowdisplaybreaks\align
  R(Z,\rho Z)&={|z+M\bar z|^2\over My^2}
               \bigg(1-{2\theta_1\theta_2\over y}\bigg)
    \equiv R_0\bigg(1-{2\theta_1\theta_2\over y}\bigg)
    \tag\NUM.\num\\    \global\plus
  r(Z,\rho Z)&={\theta_1\theta_2\over y}
               \Big[2+\chi(M^{1/2}+M^{-1/2})\Big].
  \tag\NUM.\num\endalign$$
Furthermore $j(\rho^{2k+1},Z)=\chi_\rho^{(2k+1)m}$ and
\plus
$$J^m(Z,\rho^{2k+1}Z)
  =\bigg({\zeta+2i\cosh{u\over2}\over\zeta-2i\cosh{u\over2}}\bigg)^{m/2}
  \bigg(1-{2im\chi^{2k+1}_\rho\zeta\theta_1\theta_2\over
                    y(\zeta^2+4\cosh^2{u\over2})}\bigg),
  \tag\NUM.\num$$
where $\zeta=2x\cosh{u\over2}/y$ and $u=(2k+1)\ln\sqrt{M}=(k+1/2)
l_{\rho^2}$. The evaluation of the conjugacy classes $\{\rho\}$ is
straightforward and  similar to the usual hyperbolic case. Evaluating
the relevant terms we obtain

\medskip\noindent
{\bf Theorem 4.1}:
\it
The Selberg supertrace formula for $m$-weighted Dirac-Laplace
operators $\square_m$ on compact bordered super Riemann surfaces
with Dirichlet boundary-conditions is given by:
\plus
$$\allowdisplaybreaks\align
  &\sum_{n=1}^\infty
  \Big[h(p_n^{(B)})-h(p_n^{(F)})\Big]
  =-{\hatCA\over4\pi}\int_0^\infty{g(u)-g(-u)\over\sinh{u\over2}}
  \cosh\bigg({um\over2}\bigg)\,du
  \\   &
  +\viert\sum_{\{\gamma\}}\sum_{k=1}^\infty
  {\chi_\gamma^{km}l_\gamma\over\sinh{kl_\gamma\over2}}
  \bigg[g(kl_\gamma)+g(-kl_\gamma)
  -\chi_\gamma^k\bigg(g(kl_\gamma)e^{-kl_\gamma/2}
           +g(-kl_\gamma/2)e^{kl_\gamma/2}\bigg)\bigg]
  \\   &
  -\viert\sum_{\{\rho^2\}}\sum_{k=0}^\infty
  {\chi_{\rho^2}^{(k+1/2)m}l_{\rho^2}
   \over\cosh\big[\half(k+\half)l_{\rho^2}\big]}
  \bigg\{g\big[\big(k+\bhalf\big)l_{\rho^2}\big]
        +g\big[-\big(k+\bhalf\big)l_{\rho^2}\big]
  \\  &\qquad\qquad\qquad\qquad
  -\chi_{\rho^2}^{k+\half}
   \bigg(g\big[\big(k+\bhalf\big)l_{\rho^2}\big]
                             e^{-\half(k+\half)l_{\rho^2}}
   +g\big[-\big(k+\bhalf\big)l_{\rho^2}\big]
                             e^{\half(k+\half)l_{\rho^2}}\bigg)\bigg\}
  \\   &
  -\half\sum_{i=1}^n\sum_{k=1}^\infty
   {\chi_{C_i}^{km}l_{C_i}
    \over\cosh{kl_{C_i}\over2}}\bigg[g(kl_{C_i})+g(-kl_{C_i})
  -\chi_{C_i}^k\bigg(g(kl_{C_i})e^{-kl_{C_i}/2}
           +g(-kl_{C_i})e^{kl_{C_i}/2}\bigg)\bigg],
  \\  &\quad
  \tag\NUM.\num\endalign$$
\edef\numdd{\NUM.\num}%
where $\lambda_n^{(B,F)}=\half+ip_n^{(B,F)}$ on the left runs through
the set of all eigenvalues of this Dirichlet problem, and the summation
on the right is taken over all primitive conjugacy classes
$\{\gamma\}_{\hatGamma}$, $\str(\gamma)+\chi_{\gamma}>2$,
and $\{\rho\}_{\hatGamma}$, $\rho$ hyperbolic.
\newline
The test function $h$ is required to have the following properties
\medskip
\item{i)} $h(p)\equiv h({1+m\over2}+ip)\in C^\infty(\bbbr)$,
\item{ii)} $h(p)$ need not to be an even function in $p$,
\item{iii)} $h(p)$ vanishes faster than $1/|p|$ for $p\to\pm\infty$.
\item{iv)} $h(p)$ is holomorphic in the strip $\Im(p)\leq1+{m\over2}+
          \epsilon$, $\epsilon>0$, to guarantee absolute convergence in
           the summation over $\{\gamma\}$ and $\{\rho\}$.

\medskip\noindent
Note that there is no $k=0$ contribution from the last summand.
$g(u)$ is given by Eq.(\numcm).
\newline\nobreak
\line{\hfill\vrule height0.02cm depth0.3cm width0.3cm}

\rm
\noindent
Note that in the case of Neumann boundary-conditions the last two terms
just change their signs.

\eject
\baselineskip=15pt
\noindent
{\it 2) $\hrSCF$ is non-compact.} I only consider the case $m=0$.
We now include all relevant conjugacy classes and get
\plus
$$\allowdisplaybreaks\align
  \str(\widehat L)&=
  \half\int_{\hatSCF(T)}\sum_{\{T\}}\Big[k(Z,TZ)-k(Z,T\CI Z)\Big]dV(Z)
  \\   &
  =\viert\hatCA\,\Phi(0)
  +\half\sum_{     \scriptstyle \{\gamma\}
              \atop\scriptstyle \str(\gamma)+\chi_\gamma>2         }
  \int_{\hatSCFg} k(Z,\gamma Z)
  \\   &\quad
  -\half\sum_{\{\rho\};\,\hatGamma_{\rho,hyp}}
  \int_{\hatSCF(\rho)} k(Z,\rho Z)
    \\   &\quad
  +\half\sum_{     \scriptstyle R\in\hatGamma
              \atop\scriptstyle \str(R)+\chi_R<2              }
  \int_{\hatSCF(R)} k(Z,RZ)
    \\   &\quad
  +\half\lim_{y_m\to\infty}\int_{\hySCF}dV(Z)
  \\   &\qquad\times\left\{
  \sum_{\{S\}}\sum_{\gamma'\in\hatGamma_S\backslash\Gamma}
  k(Z,{\gamma'}^{-1}S\gamma'Z)
  -\sum_{     \scriptstyle     \{\rho\};\,\hatGamma_{\rho,ell}
         \atop\scriptstyle    \str(\rho)+\chi_\rho=0}
   \sum_{\gamma'\in\hatGamma_S\backslash\Gamma}
  k(Z,{\gamma'}^{-1}\rho\gamma'Z)\right\}.
    \\   &\quad
  \tag\NUM.\num\endalign$$
\edef\numda{\NUM.\num}%
with some properly defined compact domain $\hySCF$ depending on a large
parameter $y_M$, and where the sum is taken over all hyperbolic
conjugacy classes $\{\gamma\}$, elliptic conjugacy classes $\{R\}$ and
parabolic conjugacy classes $\{S\}$ in $\bGamma$ with representatives
$\gamma$, $R$ and $S$, respectively, over all relative non-degenerate
classes $\{\rho\}$, $\rho$ hyperbolic, and over the relative conjugacy
classes $\{\rho\}$ with $\str(\rho)+\chi_\rho=0$, $\rho$ elliptic.

The hyperbolic contributions have just been calculated (c.f.\
Eq.(\numdd). Because there are no additional elliptic terms, we can
just take the result of Ref.[\GROe] [c.f.\ the discussion following
Eq.(\numbd)] and obtain
\plus
$$\multline
  \viert\int_{\hatSCF(R)} k(Z,RZ)
  =\half\sum_{\{R\}}\sum_{k=1}^{\nu-1}{1\over\nu}
  \Bigg\{\bigg(1-\chi_R^k\cos{k\pi\over\nu}\bigg)
  \\  \times
  \int_0^\infty{g(u)e^{-u/2}+g(-u)e^{u/2}\over
       \cosh u-\cos(2k\pi/\nu)}du
  +\int_0^\infty{g(u)-g(-u)\over
       \cosh u-\cos(2k\pi/\nu)}\sinh{u\over2}du\Bigg\}.
  \endmultline
  \tag\NUM.\num$$

\newpage
\baselineskip=12pt
Turning to the ``parabolic terms'' we consider the transformation
$Z\to W=\CI S^nZ$. In Ref.[\GROe] I obtained by considering $y_M$
finite with the corresponding fundamental domain $\hySCF(S)$
\plus
$$\allowdisplaybreaks\align
  &\half\int_{\hySCF(S)}{d\theta d\bar\theta\over Y} k(Z,SZ)
  \\   &
  =\half\int_0^1dx\int_0^{y_M}dy\int{d\theta d\bar\theta\over Y}
   \sum_{n\not=0}k(Z,S^nZ)
  \\   &
  =\kappa_-\Bigg\{(\ln y_M-\ln2)g(0)+\half\int_0^\infty g(-u)du
  \\   &\quad
  -{1\over4\pi}\int_{-\infty}^\infty
  \big[\Psi(1-ip)+\Psi(1+ip)\big]h(p)dp\Bigg\}
  +{\kappa\over4}\int_0^\infty\big[g(u)-g(-u)\big]du
  +O\bigg({1\over\sqrt{y_M}}\bigg)
  \\   &
  =\kappa_-\Bigg[\!\!\Bigg[(\ln y_M+C-\ln2)g(0)
  +\half\int_0^\infty g(-u)du
  \\   &\quad
  -\half\int_0^\infty\ln(1-e^{-u})
  \bigg\{{d\over du}\big[g(u)+g(-u)\big]\bigg\}du\Bigg]\!\!\Bigg]
  +{\kappa\over4}\int_0^\infty\big[g(u)-g(-u)\big]du
  +O\bigg({1\over\sqrt{y_M}}\bigg)
  \\   &\quad
  \tag\NUM.\num\endalign$$
[$\kappa_{\pm}=\sum_{\{S\}}\, (1\pm\chi_S)$],
and I have stated the result
in two alternative ways. Note that Euler's constant
$C=0.577\,215\,66490\dots$ appears only in the representation where
$g(u)$ instead of $h(p)$ is used.

As we know from the discussion in section III from the usual Selberg
case, the conjugacy class with $\tr(\rho)=0$, $\rho$ elliptic, contains
an element of order two. In the super-case this is generalized to
\plus
$$\gamma_a=\pmatrix 0 &a &0\\ -a^{-1} &0&0\\
                    0 &0 &\chi_{\gamma_a}       \endpmatrix,\qquad
  (\mod\pm1),
  \tag\NUM.\num$$
with some $a\geq1$.
Because $\gamma_a$ is an elliptic element of order two we have to
consider
\plus
$$\int_{\Cup\gamma'\hySCF,\gamma'\in\hatGamma}
  k(Z,\rho Z)
  =|\hatGamma(\rho)|
  \int_{\Cup\gamma'\hySCF,\gamma'\in
            \hatGamma\CI\backslash\hatGamma} k(Z,\rho Z)
  \tag\NUM.\num$$
and $|\hatGamma(\rho)|=\order[\hatGamma(\rho)]=2$ which
yields an additional factor $\half$ in the second term in the parabolic
contribution of Eq.(\numda).

For a proper asymptotic expansion [\VENc] of the corresponding integral
we remove from $\SCH$ two regions, denoted by $B_1^{(1|1)}=
\{Z\in\SCH|x\geq y_M\}$ and $B_2^{(1|1)}= \gamma_a B_1$, respectively,
i.e.\ we consider
\plus
$$B^{(1|1)}=\SCH-B_1^{(1|1)}-B_2^{(1|1)}.
  \tag\NUM.\num$$
First let us insert a $n=0$ ``parabolic term'' into the
super-automorphic kernel; this gives the integral
\plus
$$\int_{\Cup\gamma\hySCF,\{\gamma\}}  k(Z,\rho Z)
 =\int_{B^{(1|1)}}dV(Z) k(Z,\CI Z)+o(1),\qquad (y_m\to\infty),
  \tag\NUM.\num$$
whose asymptotic behaviour must be studied. By the definition of
$B^{(1|1)}$ the above integral separates into two contributions
\hfuzz=10pt
\plus
$$\int_0^\infty dx\int_{a^2/y_M}^{y_M}dy
 \int{d\theta_1d\theta_2\over y+\theta_1\theta_2}k(Z,\CI Z)
  -\int_0^{a^2/y_M}dy\int_{y\sqrt{a^2/yy_M-1}}^\infty dx
  \int{d\theta_1d\theta_2\over y+\theta_1\theta_2}  k(Z,\CI Z)
  \tag\NUM.\num$$
\hfuzz=1pt
In order to evaluate the first integral we set $y_0=a^2/y_M$.
\plus
$$\allowdisplaybreaks\align
  \half\int_0^\infty &dx\int_{y_0}^{y_M}dy
  \int{d\theta_1d\theta_2\over y+\theta_1\theta_2}
  \Big\{\Phi\big[R(Z,\CI Z)\big]
  +r(Z,\CI Z)\Psi\big[R(Z,\CI Z)\big]\Big\}
  \\   &
  =\int_0^\infty dx\int_{y_0}^{y_M}{dy\over y^2}
  \bigg[\half\Phi\bigg({4x^2\over y^2}\bigg)
  +{4x^2\over y^2}\Phi'\bigg({4x^2\over y^2}\bigg)
  +(1-\chi_S)\Psi\bigg({4x^2\over y^2}\bigg)\bigg]
  \\   &
  =\viert\int_0^\infty{dt\over\sqrt{t}}\int_{y_0}^{y_M}{dy\over y}
  \bigg[\half\Phi(t)+t\Phi'(t)+(1-\chi_S)\Psi(t)\bigg]
  \\   &
  =(\ln y_M-\ln a)(1-\chi_S)g(0),
  \tag\NUM.\num\endalign$$
where we have explicitly inserted $y_0$, and the terms with $\Phi$
vanish by a partial integration. Note that this term must be sufficient
for the regularization and indeed is. In the second part of the
integral of the next term we integrate out the $\theta_1\theta_2$
quantities, perform a partial integration, and get the result (set
$\tau=a^2/yy_M$)
\plus
$$\multline
  \viert\int_1^\infty{d\tau\over\tau}
  \int_{4(\tau-1)}^\infty{dt\over\sqrt{t}}
  \bigg[\half\Phi(t)+t\Phi'(t)+(1-\chi_S)\Psi(t)\bigg]
  \hfill\\   \qquad
  =\viert\int_1^\infty d(\ln\tau)
  \int_{4(\tau-1)}^\infty{dt\over\sqrt{t}}
  \bigg[\half\Phi(t)+t\Phi'(t)+(1-\chi_S)\Psi(t)\bigg]
  \hfill\\   \qquad
  =-(1-\chi_S)g(0)\ln2 +\viert\int_0^\infty{dt\over\sqrt{t}}
         \ln(t+4)\bigg[\half\Phi(t)+t\Phi'(t)+(1-\chi_S)\Psi(t)\bigg],
  \hfill\endmultline
  \tag\NUM.\num$$
and this contribution is independent of $a$ and $y_M$, respectively.
Furthermore I have used the differentiation rule
\plus
$${d\over da}\int_{\psi(a)}^{\phi(a)}f(x,a)dx
  =f[\phi(a),a]{d\phi(a)\over da}
  +f[\psi(a),a]{d\psi(a)\over da}
  +\int_{\psi(a)}^{\phi(a)}{df(x,a)\over da}dx.
  \tag\NUM.\num$$
Let us consider the various terms. Firstly we have
\plus
$$\multline
  \viert\int_0^\infty{dt\over\sqrt{t}}\ln(t+4)
  \bigg[\half\Phi(t)+t\Phi'(t)\bigg]
  \hfill\\   \qquad
  =-\viert\int_0^\infty{\sqrt{t}\,\Phi(t)\over t+4}dt
  ={1\over4\pi}\int_0^\infty{Q_1(w)\over w+4}dw
   \int_0^w{\sqrt{t}\,dt\over(t+4)\sqrt{w-t}}
  \hfill\\   \qquad
  =\half\int_0^\infty\tanh{u\over2}\tanh{u\over4}\Big[g(u)-g(-u)\Big]du.
  \hfill\endmultline
  \tag\NUM.\num$$
Here I have used the integral [\GRA, p.287]
\plus
$$\int_0^u x^{\nu-1}(x+\alpha)^\lambda(u-x)^{\mu-1}dx
   =\alpha^\lambda u^{\mu+\nu-1}B(\mu,\nu)
    {_2}F_1\bigg(-\lambda,\nu,\mu+\nu;-{u\over\alpha}\bigg)
  \tag\NUM.\num$$
together with [\EMOTa, p.101]
\plus
$$(1-z)^{1/2}{_2}F_1(a,a+\bhalf;2a;z)={_2}F_1(a-\bhalf,a;2a;z)
  =\bigg(\half+\half\sqrt{1-z}\,\bigg)^{1-2a}.
  \tag\NUM.\num$$
Next we get
\plus
$$\multline
  \viert\int_0^\infty{dt\over\sqrt{t}}\ln(t+4)\Psi(t)
  \hfill\\   \qquad
  =-{1\over4\pi}\int_0^\infty dw\,Q_2'(w)
  \int_0^w{\ln(t+4)dt\over\sqrt{t(w-t)}}
  \hfill\\   \qquad
  ={g(0)\ln2\over2}+{1\over32}\int_0^\infty{Q_2(w)dw\over
           \sqrt{1+w/4}\,\big(1+\sqrt{1+w/4}\,\big)}
  \hfill\\   \qquad
  ={g(0)\ln2\over2}+\viert\int_0^\infty
          \tanh{u\over4}\bigg\{\Big[g(u)+g(-u)\Big]
          -\tanh{u\over2}\Big[g(u)-g(-u)\Big]\bigg\}du.
  \hfill\endmultline
  \tag\NUM.\num$$
Collecting terms we obtain
\plus
$$\allowdisplaybreaks\align
  &\viert\int_1^\infty{d\tau\over\tau}
  \int_{4(\tau-1)}^\infty{dt\over\sqrt{t}}
  \bigg[\half\Phi(t)+t\Phi'(t)+(1-\chi_S)\Psi(t)\bigg]
  \\    &
  =-\half(1-\chi_S)g(0)\ln2
  \\    &\quad
  +{1-\chi_S\over4}\int_0^\infty\tanh{u\over4}\Big[g(u)+g(-u)\Big]du
  +{1+\chi_S\over4}\int_0^\infty\tanh{u\over4}
                                 \tanh{u\over2}\Big[g(u)-g(-u)\Big]du
  \\    &
  =-\half(1-\chi_S)g(0)\ln2
  \\    &\quad
  +{1-\chi_S\over4}\Bigg\{\int_0^\infty\Big[g(u)+g(-u)\Big]du
   -{2\over\pi}\int_{-\infty}^\infty h(p)
   \big[\beta(1+2ip)+\beta(1-2ip)\Big]dp\Bigg\}
  \\    &\quad
  +{1+\chi_S\over4}\Bigg\{\int_0^\infty\Big[g(u)-g(-u)\Big]du
   -{1\over\pi}\int_{-\infty}^\infty h(p)
   \bigg[\beta\bigg(\half+ip\bigg)
                    -\beta\bigg(\half-ip\bigg)\bigg]dp\Bigg\}.
  \\    &\quad
  \tag\NUM.\num\endalign$$
Here use has been made of the integrals [\GRA, p.304, p.356]
\plus
$$\int_0^\infty{e^{-\mu x}\over1+e^{-x}}dx=\beta(\mu),\qquad
  \int_0^\infty{e^{-\mu x}\over\cosh x}dx
    =\beta\bigg({\mu+1\over2}\bigg),
  \tag\NUM.\num$$
\edef\numdc{\NUM.\num}%
and $\beta(x)$ is the $\beta$-function defined by
$\beta(x)=\bhalf\big[\Psi\big(\hbox{${1+x\over2}$}\big)
         -\Psi\big(\hbox{${x\over2}$}\big)\big]$,
with $\Psi(z)=\Gamma'(z)/\Gamma(z)$ the logarithmic derivative of the
$\Gamma$-function. To finish the discussion we have to consider
\plus
$$\int_{\Cup g\gamma\hySCF,\{\gamma\}}
  k(Z,\rho Z),\qquad g\in\OSp(2|1,\bbbr).
  \tag\NUM.\num$$
According to Venkov [\VENc], this has the consequence that the
asymptotic behaviour in the limit $y_m\to\infty$ is changed by a scaling
such that the integral is calculated with respect to the variable
$\nu y_M$ instead of $y_M$. The fixed number $\nu$ is denoted by
$\nu(\rho)$. Similarly, $a$ is denoted by $a(\rho)$. Hence we obtain
\plus
$$\multline
  -\viert\int_{\hySCF}dV(Z)
  \sum_{      \scriptstyle     \{\rho\};\,\hatGamma_{\rho,ell}
         \atop\scriptstyle    \str(\rho)+\chi_\rho=0}
  k(Z,{\gamma'}^{-1}\rho\gamma'Z)
  \\   \qquad
  =-\viert q(\hatCF)(1-\chi_S)
  \left(\ln y_M-
  \sum_{    \scriptstyle     \{\rho\};\,\hatGamma_{\rho,ell}
         \atop\scriptstyle    \str(\rho)+\chi_\rho=0}
  \ln{a(\rho)\over\nu(\rho)}\right)g(0)
   +{1\over8}(1-\chi_S)q(\hatCF)g(0)\ln2
  \hfill\\    \qquad\qquad
  -{1-\chi_S\over16}q(\hatCF)
   \int_0^\infty\tanh{u\over4}\Big[g(u)+g(-u)\Big]du
  \hfill\\    \qquad\qquad
  -{1+\chi_S\over16}q(\hatCF)
   \int_0^\infty\tanh{u\over4}\tanh{u\over2}\Big[g(u)-g(-u)\Big]du+o(1).
  \hfill\endmultline
  \tag\NUM.\num$$
Here $q(\hatCF)$ denotes the number of classes $\{\rho\}$ having the
property $\str(\rho)+\chi_\rho=0$ and $\rho$ elliptic. Because we know
that all terms in the supertrace formula must be finite we deduce
$q(\hatCF)=4\kappa$. Therefore we obtain for the regularized
``parabolic terms'' in the Selberg supertrace formula for bordered
super Riemann surfaces
\hfuzz=10pt
\plus
$$\allowdisplaybreaks\align
  &\half\lim_{y_m\to\infty}\int_{\hySCF}dV(Z)
  \left\{\sum_{\{S\}}\sum_{\gamma'\in\hatGamma_S\backslash\Gamma}
  k(Z,{\gamma'}^{-1}S\gamma'Z)
  -\sum_{     \scriptstyle    \{\rho\};\,\hatGamma_{\rho,ell}
         \atop\scriptstyle    \str(\rho)+\chi_\rho=0}
   \sum_{\gamma'\in\hatGamma_S\backslash\Gamma}
  k(Z,{\gamma'}^{-1}\rho\gamma'Z)\right\}
  \\   &
  =\kappa_{S}g(0)+{\kappa_-\over2}\int_0^\infty g(-u)du
  \\   &\quad
  -{\kappa_-\over2}\int_0^\infty\ln(1-e^{-u})
   \bigg\{{d\over du}\Big[g(u)+g(-u))\Big]\bigg\}du
  +{\kappa\over4}\int_0^\infty\Big[g(u)-g(-u)\Big]du
  \\   &\quad
  -{\kappa_-\over4}\int_0^\infty\tanh{u\over4}\Big[g(u)+g(-u)\Big]du
  -{\kappa_+\over4}
   \int_0^\infty\tanh{u\over4}\tanh{u\over2}\Big[g(u)-g(-u)\Big]du.
  \tag\NUM.\num\endalign$$
\hfuzz=1pt
Here I have abbreviated
\plus
$$\kappa_{S}=\half\sum_{\{S\}}(1-\chi_S)\ln2
  -\sum_{     \scriptstyle    \{\rho\};\,\hatGamma_{\rho,ell}
         \atop\scriptstyle    \str(\rho)+\chi_\rho=0}
  \ln{a(\rho)\over\nu(\rho)}.
  \tag\NUM.\num$$
Therefore we obtain

\medskip\noindent
{\bf Theorem 4.2}:
\it
The Selberg supertrace formula for the Dirac-Laplace operator $\square$
on bordered super Riemann surfaces with hyperbolic, elliptic and
parabolic conjugacy classes with Dirichlet boundary-conditions  is
given by:
\plus
$$\allowdisplaybreaks\align
       &\sum_{n=1}^\infty
  \Big[h(p_n^{(B)})-h(p_n^{(F)})\Big]
  =i{\hatCA\over4\pi}\int_{-\infty}^\infty h(p)\tanh\pi pdp
  \\   &\quad
  +\viert\sum_{\{\gamma\}}\sum_{k=1}^\infty
  {l_\gamma\over\sinh{kl_\gamma\over2}}
  \bigg[g(kl_\gamma)+g(-kl_\gamma)
  -\chi_\gamma^k\bigg(g(kl_\gamma)e^{-kl_\gamma/2}
           +g(-kl_\gamma)e^{kl_\gamma/2}\bigg)\bigg]
  \\   &\quad
  -\viert\sum_{\{\rho^2\}}\sum_{k=0}^\infty
  {l_{\rho^2}\over\cosh\big[\half(k+\half)l_{\rho^2}\big]}
  \bigg\{g\big[\big(k+\bhalf\big)l_{\rho^2}\big]
        +g\big[-\big(k+\bhalf\big)l_{\rho^2}\big]
  \\  &\qquad\qquad\qquad\qquad
  -\chi_{\rho^2}^{k+\half}
   \bigg(g\big[\big(k+\bhalf\big)l_{\rho^2}\big]
                              e^{-\half(k+\half)l_{\rho^2}}
        +g\big[-\big(k+\bhalf\big)l_{\rho^2}\big]
                              e^{\half(k+\half)l_{\rho^2}}\bigg)\bigg\}
  \\   &\quad
  -\half\sum_{i=1}^n\sum_{k=1}^\infty
   {l_{C_i}\over\cosh{kl_{C_i}\over2}}
  \bigg\{g(kl_{C_i})+g(-kl_{C_i})
  -\chi_{C_i}^k\bigg(g(kl_{C_i})e^{-kl_{C_i}/2}
           +g(-kl_{C_i})e^{kl_{C_i}/2}\bigg)\bigg\}
  \\
  &\quad
  +\half\sum_{\{R\}}\sum_{k=1}^{\nu-1}{1\over\nu}
  \Bigg\{\bigg(1-\chi_R^k\cos{k\pi\over\nu}\bigg)
  \\   &\qquad\qquad\times
  \int_0^\infty{g(u)e^{-u/2}+g(-u)e^{u/2}\over
       \cosh u-\cos(2k\pi/\nu)}du
  +\int_0^\infty{g(u)-g(-u)\over
       \cosh u-\cos(2k\pi/\nu)}\sinh{u\over2}du\Bigg\}
  \\   &\quad
 +(\kappa_{S}+\kappa_-)g(0)+{\kappa_-\over2}\int_0^\infty g(-u)du
  \\   &\quad
  -{\kappa_-\over2}\int_0^\infty\ln(1-e^{-u})
   \bigg\{{d\over du}\Big[g(u)+g(-u)\Big]\bigg\}du
  +{\kappa\over4}\int_0^\infty\Big[g(u)-g(-u)\Big]du
  \\   &\quad
  -{\kappa_-\over4}\int_0^\infty\tanh{u\over4}\Big[g(u)+g(-u)\Big]du
  -{\kappa_+\over4}
   \int_0^\infty\tanh{u\over4}\tanh{u\over2}\Big[g(u)-g(-u)\Big]du,
  \tag\NUM.\num\endalign$$
\edef\numdb{\NUM.\num}%
where $\lambda_n^{(B,F)}=\half+ip_n^{(B,F)}$ on the left runs through
the set of all eigenvalues of this Dirichlet problem, and the summation
on the right is taken over all primitive conjugacy classes $\{\gamma\}
_{\hatGamma}$, $\str(\gamma)+\chi_\gamma>2$, $\{R\}_{\hatGamma}$,
$\str(R)+\chi_R<2$, $\{\rho\}_{\hatGamma}$, $\rho$ hyperbolic,
$\{S\}_{\hatGamma}$, $\str(S)+\chi_S=2$, and $\{\rho\}_{\hatGamma}$, $
\str(\rho)+\chi_\rho=0$, $\rho$ elliptic.
\newline
The test function $h$ is required to have the following properties
\medskip
\item{i)} $h(p)\in C^\infty(\bbbr)$,
\item{ii)} $h(p)$ need not to be an even function in $p$,
\item{iii)} $h(p)$ vanishes faster than $1/|p|$ for $p\to\pm\infty$.
\item{iv)} $h(p)$ is holomorphic in the strip $\Im(p)\leq\half+
          \epsilon$, $\epsilon>0$, to guarantee absolute convergence in
           the summation over $\{\gamma\}$ and $\{\rho\}$.
\par\nobreak
\line{\hfill\vrule height0.02cm depth0.3cm width0.3cm}

\rm
\eject\noindent
In the case of Neumann boundary-conditions the regularization procedure
is similar to the treatment in Ref.[\GROe] which is due to the fact
that in this case the continuous spectrum does not drop out and must be
taken into account. This will not discussed here again. The full
picture emerges then by a proper combination of Ref.[\GROe] and
Theorem 4.2.


\bigskip\goodbreak\noindent
\glno=0                
\advance\chapno by 1   
\line{\bf V.\ Analytic Properties of Selberg Super Zeta-Functions\hfill}
\par\medskip\nobreak\noindent
The Selberg zeta-function was originally introduced by Selberg [\SELB]
in order to study spectra of Laplacians on compact Riemann surfaces of
genus $g$. It is defined by
\plus
$$Z(s):=\prod_{\{\gamma\}}\prod_{k=0}^\infty
         \big[1-e^{-(s+k)l_\gamma}\big],\qquad(\Re(s)>1).
  \tag\NUM.\num$$
It is of further interest, because determinants of Laplacians can be
expressed by combinations of the zeta-function and its derivatives.
Define $D_\Delta(z)=\det'(-\Delta+z)$, where the prime denotes the
omission of zero modes. Then the Selberg zeta-function for compact
closed Riemann surfaces and $D_\Delta$ are connected by the relation
[\STEI, \VOROS]:
\plus
$$Z(s)=s(s-1)D_\Delta[s(s-1)]
       \big[(2\pi)^{1-s}e^{\widetilde C+s(s-1)}G(s)G(s+1)\big]^{2(g-1)},
  \tag\NUM.\num$$
where $\widetilde C=\viert-\ln\sqrt{2\pi}-2\zeta'(-1)$ and $G(z)$
denotes the Barnes $G$-function [\GRA] (see also e.g.\ [\BOSTa, \DHPHb,
\GIL]). The case of non-compact closed and open Riemann surfaces can be
found in Venkov [\VENc] (also [\BOGRO, \BOSTb]). We cite the latter
case to have a comparison with its generalization to the super-case.
The Selberg zeta-function for bordered Riemann surfaces for automorphic
$m$-forms is defined by
\plus
$$\multline
  \hat Z(s)=\prod_{\{\gamma\}}\prod_{k=0}^\infty
       \Big[1-\chi_\gamma^me^{-l_\gamma(s+k)}\Big]
  \\   \times
  \prod_{\scriptstyle \{\rho\};\,\hatGamma_{\rho,hyp}
                \atop\scriptstyle \tr(\rho)\not=0}\prod_{k=0}^\infty
  \left({1+\chi_\rho^me^{-l_\rho(s+k)}
  \over 1-\chi_\rho^me^{-l_\rho(s+k)}}\right)^{(-1)^k\chi_\CI^m}
  \times\prod_{i=1}^n\prod_{k=0}^\infty
  \left({1\over1-\chi_{C_i}^me^{-l_{C_i}(s+k)}}\right)^{2(-1)^k},
  \endmultline
  \tag\NUM.\num$$
$\Re(s)>1$. Choosing the test-function
\plus
$$h(p,s,b)={1\over(s-\half)^2+p^2}-{1\over(b-\half)^2+p^2}
  \tag\NUM.\num$$
yields

\bigskip\goodbreak\noindent
{\bf Theorem 5.1} [\BOGRO, \VENc]:
\it
The Selberg trace formula for the Selberg zeta-function on bordered
Riemann surfaces for automorphic $m$-forms is given by:
\plus
$$\allowdisplaybreaks\align
  {{\hat Z}^{\prime}(s)\over \hat Z(s)}
       &=(s-\bhalf){2\hatCA\over\pi}
  \Bigg[\Psi\bigg(s+{m\over2}\bigg)
       +\Psi\bigg(s-{m\over2}\bigg)
       -\Psi\bigg(b+{m\over2}\bigg)
       -\Psi\bigg(b-{m\over2}\bigg)\Bigg]
  \\   &\quad
  -i\sum_{\{R\}}\sum_{k=1}^{\nu-1}{1\over\nu\sin(k\pi/\nu)}
   \sum_{l=0}^\infty\Bigg[{e^{-2i(k\pi/\nu)(l+1/2-m/2)}\over s+l-m/2}
        -{e^{ 2i(k\pi/\nu)(l+1/2+m/2)}\over s+l+m/2}\Bigg]
   \\  &\quad
   +4(s-\bhalf)\sum_j\bigg({1\over(s-\half)^2+p_j^2}
                  -{1\over(b-\half)^2+p_j^2}\bigg)
   +const_1+const_2(s-\bhalf)
   \\  &\quad
   +2\tkappa\Psi(1-s)
   -4\tkappa(s-\bhalf)
    \sum_{k=0}^\infty{1\over(s-\half)^2-(k+\half)^2}
   \\  &\quad
   +\tkappa\Bigg[\Psi(s)-\Psi\bigg(s+{m\over2}\bigg)
                       -\Psi\bigg(s+{m\over2}\bigg)\Bigg],
   \tag\NUM.\num\endalign$$
with some constants $const_{1,2}$.
\newline\nobreak
\line{\hfill\vrule height0.02cm depth0.3cm width0.3cm}

\rm
\medskip\noindent
The zero- and pole-structure can be read off, see Refs.[\BOGRO, \VENc].
The functional equation has the form
\plus
$$\hat Z(1-s)=\hat Z(s)\hat\Psi_Z(s)
  \tag\NUM.\num$$
with the function $\hat\Psi_Z(s)$ given by
\plus
$$\allowdisplaybreaks\align
  \hat\Psi_Z(s)&=
  \bigg[{\Gamma(1-s)\over\Gamma(s)}\bigg]^{2\kappa}
  \exp\Bigg\{-4\hatCA\int_0^{s-\half}t
  \pmatrix \tan(\pi t)\\  \cot(\pi t)\endpmatrix dt+4\hat c(s-\bhalf)
  \\     &
  +i\sum_{\{R\}}\sum_{k=1}^{\nu-1}{1\over\nu\sin(k\pi/\nu)}
  \int_0^{s-1/2}\sum_{l=0}^\infty
   \Bigg[{e^{-2i(k\pi/\nu)(l+1/2-m/2)}\over s+l-m/2}
        +{e^{-2i(k\pi/\nu)(l+1/2-m/2)}\over s-l+(m-3)/2}
  \\     &\qquad\qquad
  -{e^{-2i(k\pi/\nu)(l+1/2+m/2)}\over s+l+(m-1)/2}
  +{e^{-2i(k\pi/\nu)(l+1/2+m/2)}\over s-l-(m-3)/2}\Bigg]
  dt\Bigg\},
  \tag\NUM.\num\endalign$$
and the $\tan(\pi t)$-, respectively the $\cot(\pi t)$-term, has to be
taken whether $m$ is even or odd, and the constant $\hat c$ given by
\plus
$$\hat c=\viert\left[
    \sum_{    \scriptstyle   \{\rho\};\,\hatGamma_{\rho,ell}
         \atop\scriptstyle   \tr(\rho)=0}\chi_\rho^m
    \ln\bigg({a(\rho)\over\mu(\rho)}\bigg)
    -\tkappa\ln2-{L\over2}\right].
  \tag\NUM.\num$$

Let us consider the two Selberg super zeta-functions $Z_0$ and $Z_1$,
respectively, defined by [\BMFSb, \GROd]
$$\allowdisplaybreaks\align
  Z_0(s)&=\prod_{\{\gamma\}}\prod_{k=0}^\infty
         \Big[1-e^{-(s+k)l_\gamma}\Big],
  \tag\NUM.\num\\    \global\plus
  Z_1(s)&=\prod_{\{\gamma\}}\prod_{k=0}^\infty
         \Big[1-\chi_\gamma e^{-(s+k)l_\gamma}\Big]
  \tag\NUM.\num\endalign$$
for $\Re(s)>1$; and furthermore the functions
$$\allowdisplaybreaks\align
  R_0(s)&={Z_0(s)\over Z_0(s+1)}=\prod_{\{\gamma\}}
               \Big(1-e^{-sl_\gamma}\Big),\qquad
  \tag\NUM.\num\\    \global\plus
  R_1(s)&={Z_1(s)\over Z_1(s+1)}=\prod_{\{\gamma\}}
         \Big(1-\chi_\gamma e^{-sl_\gamma}\Big),
  \tag\NUM.\num\endalign$$
for $\Re(s)>1$; the analytic properties of the $Z_{0,1}$-functions can
be derived from the $R_{0,1}$-functions. The analytic properties for
these functions for closed super Riemann surfaces were discussed in I
and II. However, in the case of bordered super Riemann surfaces we will
consider the {\it modified Selberg super zeta-functions} on bordered
super Riemann surfaces
$$\allowdisplaybreaks\align
  \mZO(s)&=
  \prod_{\{\gamma\}}\prod_{k=0}^\infty\Big[1-e^{-(s+k)l_\gamma}\Big]
  \\   &\qquad\times\!\!\!\!
  \prod_{     \scriptstyle\{\rho\}
         \atop\scriptstyle\str(\rho)+\chi_\rho\not=0  }
  \prod_{k=0}^\infty
  \left({1+e^{-(s+k)l_\rho}\over
         1-e^{-(s+k)l_\rho}}\right)^{(-1)^k}
  \times\prod_{i=1}^n\prod_{k=0}^\infty
  \left({1\over1-e^{-l_{C_i}(s+k)}}\right)^{2(-1)^k},
  \\ &\qquad
  \tag\NUM.\num
  \\   \global\plus
  \mZE(s)&=
  \prod_{\{\gamma\}}\prod_{k=0}^\infty
         \Big[1-\chi_\gamma e^{-(s+k)l_\gamma}\Big]
  \\   &\qquad
  \times\!\!\!\!
  \prod_{     \scriptstyle\{\rho\}
         \atop\scriptstyle\str(\rho)+\chi_\rho\not=0  }
  \prod_{k=0}^\infty
  \left({1+\chi_\rho e^{-(s+k)l_\rho}
        \over
     1-\chi_\rho e^{-(s+k)l_\rho}}\right)^{(-1)^k}
  \times\prod_{i=1}^n\prod_{k=0}^\infty
  \left({1\over1-\chi_{C_i}e^{-l_{C_i}(s+k)}}\right)^{2(-1)^k}
  \\ &\qquad
  \tag\NUM.\num\endalign$$
for $\Re(s)>1$. For convenience we will consider the functions
$$\allowdisplaybreaks\align
  &\mRO(s):={\mZO(s)\over\mZO(s+1)}
  =\prod_{\{\gamma\}}\Big(1-e^{-sl_\gamma}\Big)
  \\   &\qquad\times\!\!\!\!
  \prod_{     \scriptstyle\{\rho\}
         \atop\scriptstyle\str(\rho)+\chi_\rho\not=0  }
  \!\!\!\!\prod_{k=0}^\infty
  \left({1+e^{-(s+k)l_\rho}\over
         1-e^{-(s+k)l_\rho}}\right)^{\alpha_k(-1)^k}
  \!\!\!\!\times\prod_{i=1}^n\prod_{k=0}^\infty
  \left({1\over1-e^{-l_{C_i}(s+k)}}\right)^{2\alpha_k(-1)^k},
  \\ &\qquad
  \tag\NUM.\num\\  \global\plus
  &\mRE(s):={\mZE(s)\over\mZE(s+1)}=
  \prod_{\{\gamma\}}\Big(1-\chi_\gamma e^{-sl_\gamma}\Big)
  \\   &\qquad\times\!\!\!\!
  \prod_{     \scriptstyle\{\rho\}
         \atop\scriptstyle\str(\rho)+\chi_\rho\not=0  }
  \!\!\!\!\prod_{k=0}^\infty
  \left({1+\chi_\rho e^{-(s+k)l_\rho}\over
  1-\chi_\rho e^{-(s+k)l_\rho}}\right)^{\alpha_k(-1)^k}
  \!\!\!\!\times\prod_{i=1}^n\prod_{k=0}^\infty
  \left({1\over1-\chi_{C_i}e^{-l_{C_i}(s+k)}}\right)^{2\alpha_k(-1)^k},
  \\ &\qquad
  \tag\NUM.\num\endalign$$
\advance\glno by -1
\edef\numeb{\NUM.\num}\plus
\edef\numea{\NUM.\num}%
$\alpha_k=1\,(m=0)$, $\alpha_k=2\,(k\in\bbbn)$, and $\Re(s)>1$. As we
shall see, only functional relations for the ${\hat R}_{0,1}$-functions
can be derived, but not for the ${\hat Z}_{0,1}$-functions.

\medskip\noindent
{\it 1.\ The Selberg super zeta-function $R_1$}.
We first discuss the function $\mRE(s)$. In order to do this we choose
the test function [\GROd] $(\Re(s,a)>1)$
\plus
$$h_1(p,s,a)=2ip\bigg({1\over s^2+p^2}-{1\over a^2+p^2}\bigg),
  \tag\NUM.\num$$
with the Fourier transformed function $g_1(u)$ given by
\plus
$$g_1(u,s,a)=\sign(u)\Big(e^{-s|u|}-e^{-a|u|}\Big).
  \tag\NUM.\num$$
The regularization term is needed to match the requirements of a valid
test function in the trace formula. The relevant integrals have been
already calculated in Refs.[\GROd, \GROe], such that we just take the
results. The exceptions are the new terms in the supertrace formula
corresponding to the involuted orbits and the last summand in
Eq.(\numdb). For the latter we obtain by using the integrals (\numdc)
\plus
$$\allowdisplaybreaks\align
  \int_0^\infty
  &\tanh{u\over4}\tanh{u\over2}\Big[g_1(u,s,a)-g_1(-u,s,a)\Big]du
  \\   &=4\int_0^\infty\bigg(1-{1\over\cosh u}\bigg)
    \bigg(e^{-2su}-e^{-2au}\bigg)du
  \\   &=2\bigg[{1\over s}-{1\over
                     a}-\Psi\bigg({s\over2}+{3\over4}\bigg)
  +\Psi\bigg({s\over2}+\viert\bigg)
  +\Psi\bigg({a\over2}+{3\over4}\bigg)
  -\Psi\bigg({a\over2}+\viert\bigg)\bigg].
  \tag\NUM.\num\endalign$$
Let us consider the term corresponding to the $\{\rho\}$-conjugacy
classes in Eq.(\numdb). We get by inserting $g_1(u)$ on the one hand
\plus
$$\multline
    -\viert\sum_{\{\rho^2\}}\sum_{k=0}^\infty
  {l_{\rho^2}\over\cosh\big[\half(k+\half)l_{\rho^2}\big]}
  \bigg\{g\big[\big(k+\bhalf\big)l_{\rho^2}\big]
        +g\big[-\big(k+\bhalf\big)l_{\rho^2}\big]
        \\  \qquad
  -\chi_{\rho^2}^{k+\half}
   \bigg(g\big[\big(k+\bhalf\big)l_{\rho^2}\big]
                              e^{-\half(k+\half)l_{\rho^2}}
        +g\big[-\big(k+\bhalf\big)l_{\rho^2}\big]
                              e^{\half(k+\half)l_{\rho^2}}\bigg)\bigg\}
  \hfill\\  \qquad
  =-\half\sum_{\{\rho^2\}}l_{\rho^2}\sum_{k=0}^\infty
  \chi_{\rho^2}^{k+\half}\tanh[\bhalf(k+\bhalf)l_{\rho^2}]
  \Big(e^{-s(k+\half)l_{\rho^2}}-e^{-a(k+\half)l_{\rho^2}}\Big).
  \hfill\endmultline
  \tag\NUM.\num$$
On the other we have by means of
\plus
$$\ln{1+x\over1-x}=2\sum_{k=0}^\infty{x^{2k+1}\over2k+1}
  \tag\NUM.\num$$
for the logarithmic derivative of the inverse hyperbolic terms of
the $\mRE$-function
\plus
$$\allowdisplaybreaks\align
  {d\over ds}&\ln\prod_{\{\rho\}}\prod_{m=0}^\infty
  \left({1+\chi_\rho e^{-(s+m)l_\rho}\over
  1-\chi_\rho e^{-(s+m)l_\rho}}\right)^{\alpha_m(-1)^m}
  \\    &
  =2\sum_{\{\rho\}}\sum_{k=0}^\infty
   {\chi_\rho^{2k+1}\over2k+1}\bigg({d\over ds}e^{-sl_\rho(2k+1)}\bigg)
   \sum_{m=0}^\infty\alpha_m(-1)^me^{-ml_\rho(2k+1)}
  \\    &
  =-\sum_{\{\rho^2\}}l_{\rho^2}\sum_{k=0}^\infty
   \chi_{\rho^2}^{k+\half}e^{-sl_{\rho^2}(k+\half)}
   \left[1+2\sum_{m=1}^\infty(-1)^me^{-ml_{\rho^2}(k+\half)}\right]
  \\    &
  =-\sum_{\{\rho^2\}}l_{\rho^2}\sum_{k=0}^\infty
   \chi_{\rho^2}^{k+\half}e^{-sl_{\rho^2}(k+\half)}
   \tanh\big[\bhalf l_{\rho^2}(k+\bhalf)\big].
  \tag\NUM.\num\endalign$$
And similarly for the third term in $\mRE$. Therefore we  obtain the
Selberg supertrace formula for the test function $h_1(p,s,a)$ as follows
\plus
$$\multline
  \!\!\!\!
  {\mREp(s)\over\mRE(s)}-{\mREp(a)\over\mRE(a)}
  \\
  =4\sum_{n=1}^\infty\left[
  {\lambda_n^{(B)}-\half\over s^2-(\lambda_n^{(B)}-\half)^2}
  -{\lambda_n^{(B)}-\half\over a^2-(\lambda_n^{(B)}-\half)^2}
  -{\lambda_n^{(F)}-\half\over s^2-(\lambda_n^{(F)}-\half)^2}
  +{\lambda_n^{(F)}-\half\over a^2-(\lambda_n^{(F)}-\half)^2}\right]
  \hfill\\   \qquad
  -2\sum_{\{R\}}\sum_{k=1}^{\nu-1}{\chi_R^k\over\nu}
  \sum_{l=0}^\infty\cos\bigg[(2l+1){k\pi\over\nu}\bigg]
   \bigg({1\over s+l+\half}-{1\over a+l+\half}\bigg)
  \hfill\\  \qquad
  -{\hatCA\over2\pi}\Big[\Psi(s+\bhalf)-\Psi(a+\bhalf)\Big]
  +\kappa\bigg({1\over s}-{1\over a}\bigg)
  \hfill\\  \qquad
  -\kappa_+\bigg[\Psi\bigg({s\over2}+{3\over4}\bigg)
  -\Psi\bigg({s\over2}+\viert\bigg)-\Psi\bigg({a\over2}+{3\over4}\bigg)
  +\Psi\bigg({a\over2}+\viert\bigg)\bigg].
  \hfill\endmultline
  \tag\NUM.\num$$
\edef\numef{\NUM.\num}%
Thus we read off

\medskip\noindent
{\bf Theorem 5.2}:
\it
The Selberg super zeta-function $\mRE(s)$ is a meromorphic function on
$\Li$ and has furthermore the following properties:
\medskip
\item{A)}
 The Selberg super zeta-function $\mRE(s)$ has
``trivial'' zeros at the following points and nowhere else
\itemitem{i)} $s=-\half-l$, $(l=0,1,2,\dots)$
          and the multiplicity of these zeros is given by
\plus
$$\# N_l={\hatCA\over2\pi}
         -2\sum_{\{R\}}\sum_{k=1}^{\nu-1}{\chi_R^k\over\nu}
  \sum_{l=0}^\infty\cos\bigg[(2l+1){k\pi\over\nu}\bigg].
  \tag\NUM.\num$$
Note that if $\# N_l<0$, we have poles instead of zeros.
\itemitem{ii)} $s=0$ with the multiplicity given by
           $\# N_0=\kappa$.
\itemitem{iii)} $s=-{3\over2}-2l$, $(l=0,1,2,\dots)$, with the
            multiplicity given by $\# N_l=2\kappa_+$.
\item{B)}
 The Selberg super zeta-function $\mRE(s)$ has
``trivial'' poles at the following points and nowhere else
\itemitem{i)} $s=-\half-2l$ $l=0,-1,-2,\dots$ with the multiplicity
          given by $\# P_l=2\kappa_+$.
\item{C)}
 The Selberg super zeta-function $\mRE(s)$ has ``non-trivial'' zeros
and poles at the following points and nowhere else [\GROd]
\itemitem{i)} $s=ip_n^{(B,F)}$: there are zeros (poles) with twice the
          multiplicity as the corresponding eigenvalue of $\square$.
\itemitem{ii)} $s=-ip_n^{(B,F)}$: reversed situation for poles and
          zeros.
\itemitem{iii)} $s=\lambda^{(B,F)}_n-\half$ there are zeros (poles), and
\itemitem{iv)} $s=-(\lambda^{(B,F)}_n-\half)$ there are poles with twice
           the multiplicity as the corresponding eigenvalue of
           $\square$, respectively. The last two cases describe
           so-called small eigenvalues of the operator $\square$. All
           these ``nontrivial'' eigenvalues are supernumbers $s\in\Li$.

\medskip
\noindent
Of course, Eq.(\numef) can be extended meromorphically
to all $s\in\Li$.
\newline
\line{\hfill\vrule height0.02cm depth0.3cm width0.3cm}

\rm
\medskip\noindent
By means of the relation (\numea) the analytic properties of the Selberg
super zeta-functions $\mZE$ can be derived, compare also Ref.[\GROd].

The test functions $h_1(p,s,a)$ is symmetric by the interchange
$s\to-s$. Therefore subtracting the trace formula for $h_1(p,s,a)$ and
$h_1(p,-s,a)$ yields the functional equation for $\mRE$ in differential
form
\plus
$$\multline
  {d\over ds}\ln\big[\mRE(s)\mRE(-s)\big]
  \\   \qquad
  =-{\hatCA\over2}\tan\pi s
  -\kappa_+\bigg[\Psi\bigg({s\over2}+{3\over4}\bigg)
  -\Psi\bigg({3\over4}-{s\over2}\bigg)-\Psi\bigg(\viert+{s\over2}\bigg)
  +\Psi\bigg(\viert-{s\over2}\bigg)\bigg]
  \hfill\\
  +{2\kappa\over s}  -2\sum_{\{R\}}\sum_{k=1}^{\nu-1}{\chi_R^k\over\nu}
  \sum_{l=0}^\infty\cos\bigg[(2l+1){k\pi\over\nu}\bigg]
   \bigg({1\over s+l+\half}+{1\over s-(l+\half)}\bigg)
  \endmultline
  \tag\NUM.\num$$
[note $\Psi(\half+s)=\Psi(\half-s)+\pi\tan\pi s$].
The integrated functional equation therefore has the form
\plus
$$\mRE(s)\mRE(-s)=const.\,(\cos\pi s)^{\hatCA/2\pi}  s^{2\kappa}
  \left[{\Gamma(\viert+{s\over2})\Gamma(\viert-{s\over2})
  \over \Gamma({3\over4}-{s\over2})\Gamma({3\over4}+{s\over2})}\right]
  ^{2\kappa_+}\mPsiE(s),
  \tag\NUM.\num$$
with the function $\mPsiE(s)$ given by
\plus
$$\mPsiE(s)=\exp\left\{
  -2\sum_{\{R\}}\sum_{k=1}^{\nu-1}{\chi_R^k\over\nu}
  \sum_{l=0}^\infty\cos\bigg[(2l+1){k\pi\over\nu}\bigg]
  \ln\Big|s^2-(l+\bhalf)^2\Big|\right\}.
  \tag\NUM.\num$$
We can check the consistence of the functional equation with respect to
the analytical properties of the Selberg super-zeta function $\mRE$. In
the case that there are no elliptical and parabolic terms the
functional equation simplifies into
\plus
$$\mRE(s)\mRE(-s)=A_1(\cos\pi s)^{\hatCA/2\pi},
  \tag\NUM.\num$$
where $A_1$ is a constant given e.g.\,by $A_1=R_1(s_0)R_1(-s_0)(\cos\pi
s_0)^{-\hatCA/2\pi}$ with some $s_0\in\bbbc$, which is however,
independent of $s_0$.

\medskip\noindent
{\it 2.\ The Selberg super zeta-function $\mRO$.}
Let us turn to the discussion of the Selberg super zeta-function $\mRO$.
We consider the test-function $(\Re(s,a)>1)$
\plus
$$h_0(p,s,a)=2\Big(\bhalf+ip\Big)\bigg(
  {1\over s^2-(\half+ip)^2}-{1\over a^2-(\half+ip)^2}\bigg),
  \tag\NUM.\num$$
with the Fourier transform $g_0(u,s,a)$ given by
\plus
$$g_0(u,s,a)=\sign(u)e^{u/2}\Big(e^{-s|u|}-e^{-a|u|}\Big).
  \tag\NUM.\num$$
Again the regularization term is needed to match the requirements of a
valid test function for the trace formula. Similarly as in the previous
case we obtain the Selberg super trace formula for the test function
$h_0(p,s,a)$ as follows
\hfuzz=15pt
\plus
$$\allowdisplaybreaks\align
  &{\mROp(s)\over\mRO(s)}-{\mROp(a)\over\mRO(a)}
  \\  &
  =4\sum_{n=1}^\infty\left[
  {\lambda_n^{(B)}\over s^2-(\lambda_n^{(B)})^2}
  -{\lambda_n^{(B)}\over a^2-(\lambda_n^{(B)})^2}
  -{\lambda_n^{(F)}\over s^2-(\lambda_n^{(F)})^2}
  +{\lambda_n^{(F)}\over a^2-(\lambda_n^{(F)})^2}\right]
  \\   &\qquad
  -\sum_{\{R\}}\sum_{k=1}^{\nu-1}  {1\over\nu\sin(2k\pi/\nu)}
  \sum_{l=1}^\infty\sin\bigg({2lk\pi\over\nu}\bigg)
  \left[{1\over s+l-1}-{1\over s+l+1}-{1\over a+l-1}+{1\over a+l+1}\right]
  \\   &\qquad
  -{\hatCA\over4\pi}\Big[\Psi(s)+\Psi(s+1)-\Psi(a)-\Psi(a+1)\Big]
  \\   &\qquad
  +{\kappa\over2}\bigg({1\over s-\half}+{1\over s+\half}
    -{1\over a-\half}-{1\over a+\half}-{4\over s}+{4\over a}\bigg).
  \tag\NUM.\num\endalign$$
\edef\numee{\NUM.\num}%
\hfuzz=1pt
Note that no terms proportional to $\chi$ are present. Therefore we
have shown

\medskip\noindent
{\bf Theorem 5.3}:
\it
The Selberg super zeta-function $\mRO(s)$ is a meromorphic function on
$\Li$ and has furthermore the following properties:
\medskip
\item{A)}
 The Selberg super zeta-function $\mRO(s)$ has ``trivial'' zeros at the
following points and nowhere else: First note that
$${1\over\sin(2k\pi/\nu)}\sum_{l=1}^\infty
  \sin\bigg({2lk\pi\over\nu}\bigg)
  \bigg({1\over s+l-1}-{1\over s+l+1}\bigg)
  ={1\over s}  +{\cos({2k\pi\over\nu})\over s+1}
  +2\sum_{l=2}^\infty{\cos({2lk\pi\over\nu})\over s+l}.$$
\itemitem{i)} $s=0$ with multiplicity
\plus
$$\# N_0={\hatCA\over4\pi}-2\kappa -\sum_{\{R\}}{\nu-1\over\nu}.
  \tag\NUM.\num$$
$s=-1$ with multiplicity
\plus
$$\# N_1={\hatCA\over2\pi}-\sum_{\{R\}}\sum_{k=1}^{\nu-1}
  {1\over\nu}\cos\bigg({2k\pi\over\nu}\bigg).
  \tag\NUM.\num$$
$s=-n$ $(n=2,3,\dots)$ with multiplicity
\plus
$$\# N_n={\hatCA\over2\pi}-2\sum_{\{R\}}\sum_{k=2}^{\nu-1}
  {1\over\nu}\sum_{l=2}^\infty\cos\bigg({2lk\pi\over\nu}\bigg).
  \tag\NUM.\num$$
Note that if $\# N_n<0$, we have poles instead of zeros.
\itemitem{ii)} $s=-\half$ with multiplicity $\# N_{-1/2}=\kappa/2$.
\itemitem{iii)} $s=\half$ with multiplicity $\# N_{\half}=\kappa/2$.
\item{B)}
 The Selberg super zeta-function $\mRO(s)$ has ``non-trivial'' zeros
and poles at the following points and nowhere else [\GROd]
\itemitem{i)} $s=ip_n^{(B,F)}+\half$: there are zeros (poles) with twice
          the multiplicity as the corresponding eigenvalue of $\square$.
\itemitem{ii)} $s=-ip_n^{(B,F)}-\half$:
           reversed situation for poles and zeros.
\itemitem{iii)} $s=\lambda^{(B,F)}_n$ there are zeros (poles), and
\itemitem{iv)} $s=-\lambda^{(B,F)}_n$ there are poles with twice the
           multiplicity as the corresponding eigenvalue of $\square$,
           respectively. The last two cases describe so-called small
           eigenvalues of the operator $\square$. All these
           ``nontrivial'' eigenvalues are supernumbers $s\in\Li$.

\medskip\noindent
Of course, Eq.(\numee) can be extended meromorphically
to all $s\in\Li$.
\newline
\line{\hfill\vrule height0.02cm depth0.3cm width0.3cm}

\rm
\medskip\noindent
By means of the relation (\numeb) the analytic properties of the Selberg
super zeta-functions $\mZO$ can be derived, compare also Ref.[\GROd].

The test function $h_0(p,s,a)$ is symmetric with respect to
$s\to-s$. Therefore subtracting the trace formul\ae\ of
$h_0(p,s,a)$ and $h_0(p,-s,a)$ from each other yields the
functional equation for the $\mRO$-function in differential form
\plus
$$\multline
  {d\over ds}\ln\Big[\mRO(s)\mRO(-s)\Big]
  \\   \qquad
  ={\hatCA\over2\pi}{d\over ds}\ln(\sin\pi s)
  +\kappa\bigg({1\over s+\half}+{1\over s-\half}-{4\over s}\bigg)
  \hfill\\  \qquad\qquad
  -\sum_{\{R\}}\sum_{k=1}^{\nu-1}
  {1\over\nu\sin(2k\pi/\nu)}
  \sum_{l=1}^\infty\sin\bigg({2lk\pi\over\nu}\bigg)
  \hfill\\  \times
  \left[{1\over s+l-1}+{1\over s-(l-1)}-{1\over s+l+1}-{1\over s-(l+1)}
  \right].
  \endmultline
  \tag\NUM.\num$$
In integrated form, this gives the functional equation
\plus
$$\mRO(s)\mRO(-s) =  const.\,(\sin\pi s)^{\hatCA/2\pi}
  \bigg({s^2-\viert\over s^4}\bigg)^{\kappa}\,\mPsiO(s),
  \tag\NUM.\num$$
with the function $\mPsiO(s)$ given by
\plus
$$\mPsiO(s)=\exp\left\{-\sum_{\{R\}}\sum_{k=1}^{\nu-1}
  {1\over\nu\sin(2k\pi/\nu)}
  \sum_{l=1}^\infty\sin\bigg({2lk\pi\over\nu}\bigg)
  \ln\left|{(s^2-(l-1)^2)\over(s^2-(l+1)^2)}\right|\right\}.
  \tag\NUM.\num$$
We can check the consistence of the functional equation with respect to
the analytical properties of the Selberg super-zeta function $\mRO$.
In the case that there are no elliptical and parabolic terms the
functional equation simplifies into
\plus
$${\mZO(s)\mZO(-s)\over\mZO(1+s)\mZO(1-s)}=
  \mRO(s)\mRO(-s)=A_0(\sin\pi s)^{\hatCA/2\pi},
  \tag\NUM.\num$$
where the constant $A_0$ is e.g.\,given by $A_0
=\mRO(s_0)\mRO(-s_0) (\sin\pi s_0)^{-\hatCA/2\pi}$ with some
$s_0\in\bbbc$, where $\widetilde B_0$ is independent of $s_0$.

\medskip\noindent
{\it 3.\ The Selberg super zeta-function $\mZS$.}
Following Refs.[\GROd, \GROe, \MUYa] we can also introduce the Selberg
super zeta-function $\mZS(s)$ defined by
\plus
$$\mZS(s)={\mZO(s)\mZO(s+1)\over\mZEz(s+\half)}.
  \tag\NUM.\num$$
The appropriate test function is $(\Re(s)>1)$
\plus
$$h_S(p,s)={1\over s^2-\lambda^2}\bigg|_{\lambda=\half+ip}
  ={1\over (s^2-\viert)-ip+p^2}.
  \tag\NUM.\num$$
The corresponding Fourier transform $g_S$ is given by
\plus
$$g_S(u,s)={1\over2s}e^{u/2-s|u|}.
  \tag\NUM.\num$$
The evaluation of the various terms in the Selberg supertrace
formula is straightforward similarly to the previous two cases.
We just present the evaluation of the fourth last term in Eq.(\numdb)
proportional to $\kappa_-$. We obtain
\plus
$$\allowdisplaybreaks\align
  \int_0^\infty&\ln(1-e^{-u})
   \bigg\{{d\over du}\Big[g(u)+g(-u))\Big]\bigg\}du
  \\    &
  =-{1\over2s}\int_0^\infty\ln(1-e^{-u})
    \Big[(s-\bhalf)e^{-u(s-\half)}
             +(s+\bhalf)e^{-u(s+\half)}\Big]
  \\    &
  =-{1\over2s}\int_0^1\ln x
   \Big[(s+\bhalf)(1-x)^{s-\half}+(s-\bhalf)(1-x)^{s-{3\over2}}\Big]
  \\    &
  ={1\over2s}\Big[2C+\Psi(s+\hbox{${3\over2}$})+\Psi(s+\bhalf)\Big].
  \tag\NUM.\num\endalign$$
Here use has been made of the integral [\GRA, p.538]
\plus
$$\int_0^1 x^{\mu-1}(1-x^r)^{\nu-1}\ln x dx
  ={1\over r^2}B\bigg({\mu\over r},\nu\bigg)
   \bigg[\Psi\bigg({\mu\over r}\bigg)
        -\Psi\bigg({\mu\over r}+\nu\bigg)\bigg].
  \tag\NUM.\num$$
Therefore we obtain
\plus
$$\allowdisplaybreaks\align
  &{1\over 2s}{\mZSp(s)\over\mZS(s)}
  ={1\over 2s}{d\over ds}\ln\bigg[{\mZO(s)\mZO(s+1)\over\mZEz(s+\half)}\bigg]
  \\    &=
  2\sum_{n=1}^\infty\left[{1\over s^2-(\lambda_n^{(B)})^2}
                   -{1\over s^2-(\lambda_n^{(F)})^2}\right]
  +{\hatCA\over8\pi}{1\over s^2}
  \\   &\quad
  -{1\over2s}\sum_{\{R\}}\sum_{k=1}^{\nu-1}
  \sum_{l=1}^\infty
  {\sin(2lk\pi/\nu)\over\nu\sin(2k\pi/\nu)}
  \bigg[{4(1-\chi_R^k\cos({k\pi\over\nu}))\over s+l}
  +{1\over s+l-1}+{1\over s+l+1}-{2\over s+l}\bigg]
  \\   &\quad
  +{C\kappa_{-}-\kappa_s-\kappa_-\over s}
  +{\kappa\over4s}
   \bigg({4\over s}+{1\over s-\half}-{1\over s+\half}\bigg)
  \\   &\quad
  +{\kappa_-\over s}\Psi(s)
  +{\kappa_+\over2s}\bigg[\Psi\bigg({s\over2}\bigg)
             -\Psi\bigg({s+1\over2}\bigg)\bigg].
  \tag\NUM.\num\endalign$$
\edef\numed{\NUM.\num}%
This gives

\medskip\noindent
{\bf Theorem 5.4}:
\it
The Selberg super zeta-function $\mZS$ is a meromorphic function on
$\Li$ and has furthermore the following properties:
\newline
A) The Selberg super zeta-function $\mZS(s)$ has
``trivial'' zeros at the following points and nowhere else
\medskip
\item{i)} $s=0$ with multiplicity
\plus
$$\# N_0={\hatCA\over4\pi}-\sum_{\{R\}}{\nu-1\over\nu}+2\kappa.
  \tag\NUM.\num$$
{\it $s=-1$ with multiplicity}
\plus
$$\# N_1=-2\sum_{\{R\}}\sum_{k=1}^{\nu-1}{1\over\nu}
  \bigg[1-2\chi_R\cos\bigg({k\pi\over\nu}\bigg)
          +\cos\bigg({k\pi\over\nu}\bigg)\bigg]-4\kappa.
  \tag\NUM.\num$$
$s=-n$ $(n=-2,3,4,\dots)$ with multiplicity
\plus
$$\# N_n=4\sum_{\{R\}}\sum_{k=2}^{\nu-1}
   {1\over\nu\sin(2k\pi/\nu)}
   \bigg[\sin^2\bigg({k\pi\over\nu}\bigg)
   -\bigg(1-\chi_R\cos{k\pi\over\nu}\bigg)\bigg]
   \sin\bigg({2lk\pi\over\nu}\bigg)-4\kappa.
  \tag\NUM.\num$$
\it
\item{ii)} $s=\half$ with multiplicity $\# N_{1/2}=\kappa/2$.
\item{iii)} $s=-1-2l$, $l=0,1,2,\dots$, with multiplicity
            $\# N_l=2\kappa_+$.

\bigskip\nobreak\noindent
Note that if $\# N_l<0$, we have poles instead of zeros.

\bigskip\goodbreak\noindent
B) The Selberg super zeta-function $\mZS(s)$ has
``trivial'' poles at the following points and nowhere else
\item{i)} $s=-\half$ with multiplicity $\# P_{-1/2}=\kappa/2$.

\noindent
C) The Selberg super zeta-function $\mZS(s)$ has ``non-trivial'' zeros
and poles at the following points and nowhere else [\GROd, \MUYa]
\item{i)} $s=\pm(\half+ip_n^{(B)})$ there are zeros and
\item{ii)} $s=\pm(\half+ip_n^{(F)})$ there are poles, with twice the
           multiplicity as the corresponding eigenvalue of
           $\square$, respectively.

\bigskip\goodbreak\noindent
Of course, Eq.(\numed) can be extended meromorphically
to all $s\in\Li$.
\newline\nobreak
\line{\hfill\vrule height0.02cm depth0.3cm width0.3cm}

\rm
\medskip\goodbreak\noindent
The test function $h_S(p,s)$ is symmetric with respect
to $s\to-s$ and therefore we can deduce the functional relation
\plus
$$\multline
  {\mZS(s)\over\mZS(-s)}
  =const.\,e^{4s(C\kappa_{-}-\kappa_{S}-\kappa_-)}
  \\   \times
   \bigg({s-\half\over s+\half}\bigg)^{\kappa}
   \bigg({\Gamma(s)\over\Gamma(-s)}\bigg)^{2\kappa_-}
   \bigg({\Gamma({s\over2})\Gamma(\half-{s\over2})\over
          \Gamma(\half-{s\over2})\Gamma(-{s\over2})}\bigg)^{2\kappa_+}
  \mPsiS(s),
  \endmultline
  \tag\NUM.\num$$
\edef\numeg{\NUM.\num}%
with the function $\mPsiS(s)$ given by
\plus
$$\multline
  \mPsiS(s)=\exp\left\{-2\sum_{\{R\}}\sum_{k=1}^{\nu-1}
  {1\over\nu\sin(2k\pi/\nu)}
  \sum_{l=1}^\infty\sin\bigg({2lk\pi\over\nu}\bigg)
  \right.\\   \times\left.\vphantom{\sum_{\{R\}}^1}
  \Bigg[2\bigg(1-2\chi_R^k\cos{k\pi\over\nu}\bigg)
      \ln\bigg|{s+l\over s-l}\bigg|
      +\ln\bigg|{(s+l-1)(s+l+1)\over(s-l+1)(s-l-1)}\bigg|\Bigg]\right\}.
  \endmultline
  \tag\NUM.\num$$
We can check the consistence of the functional equation with
respect to the analytical properties of the Selberg super-zeta
function $\mZS$. In the case, where only hyperbolic conjugacy classes
are present in the super Fuchsian group, Eq.(\numeg) reduces to the
simple functional equation [\GROd]
\plus
$$\mZS(s)=\mZS(-s).
  \tag\NUM.\num$$
Let us note that the relation
\plus
$$\multline
  {d\over ds}\ln\bigg[{\mZO(s)\mZO(s+1)\over\mZE(s+\half)}\bigg]
  -{d\over ds}\ln\bigg[{\mZO(s+1)\mZO(s+2)\over\mZE(s+{3\over2})}\bigg]
  \\
  ={\mROp(s)\over\mRO(s)}+{\mROp(s+1)\over\mRO(s+1)}
  -2{\mREp(s+\half)\over\mRE(s+\half)}
  \endmultline
  \tag\NUM.\num$$
provides a consistency check for the zeta functions $\mRO$, $\mRE$ and
$\mZS$, respectively.

Let us note that in the case of Neumann boundary-conditions the Selberg
super zeta-functions must be differently defined due to the changed
signs of the $\gamma\CI$-terms, i.e.\ the power of the corresponding
terms in the zeta-functions is reversed. This concludes the discussion.


\bigskip\goodbreak\noindent
\glno=0                
\advance\chapno by 1   
\line{\bf VI.\ Super-Determinants of Dirac-Laplace Operators\hfill}
\par\medskip\nobreak\noindent
Since $\square_m^2$ is not a positive definite operator I calculate the
super-determinants of $(c^2-\square_m^2)$ for $\Re(c)>0$ and
analytically continue in $c$. Similar considerations have been done by
Aoki [\AOK] and Ref.[\GROd] by means of the supertrace of the heat
kernel of $\square_m^2$ for the case of closed super Riemann surfaces.

The super-determinant is defined using the $\zeta$-function
regularization in the following way
\plus
$$\allowdisplaybreaks\align
  \sdet(c^2-\square_m^2)&=\exp\left[-{\partial\over\partial s}
                        \zeta(s;c)\bigg|_{s=0}\right]
  \\
  \zeta(s;c)          &=\str\big[(c^2-\square_m^2)^{-s}\big]
  ={1\over\Gamma(s)}\int_0^\infty dt\,t^{s-1}
                          \str\{\exp[-t(c^2-\square_m^2)]\}.
  \tag\NUM.\num\endalign$$
\edef\numeh{\NUM.\num}%
The function $h$ corresponding to the heat-kernel of $(c^2-\square_m^2)$
reads $h_{hk}(s)=e^{t[(s+m/2)^2-c^2]}$. This gives
\plus
$$G(u,\chi)={1\over\sqrt{\pi t}}e^{-u^2/4t-c^2t}
  \bigg[\cosh(m+1){u\over2}-\chi\cosh{um\over2}\bigg].
  \tag\NUM.\num$$
Splitting now the calculation of $\zeta(s;c)$ into two terms
corresponding to the identity transformation and the length term,
respectively, gives:
\plus
$$\zeta(s;c)
  =\zeta_I(s;c)+\zeta_\Gamma(s;c).
  \tag\NUM.\num$$
We have [\AOK, \GROd]
\plus
$$\zeta_I(s;c)
  =-{\hatCA\over8\pi\Gamma(s)}\sum_{k=0}^m
  \int_0^\infty t^{s-1}e^{-(c^2-k^2)t}=
  -{\hatCA\over8\pi}\sum_{k=0}^m(c^2-k^2)^{-s},
  \tag\NUM.\num$$
therefore
\plus
$${\partial\over\partial s}\zeta_I(s;c)\bigg|_{s=0}=
  {\hatCA\over8\pi}\sum_{k=0}^m\ln(c^2-k^2).
  \tag\NUM.\num$$
Let us consider the representation ($\Re(s)<1$):
\plus
$$t^{s-1}={2\over\Gamma(1-s)}\int_0^\infty
  {\lambda+c\over[\lambda(\lambda+2c)]^s}
  e^{-\lambda(\lambda+2c)t}d\lambda;
  \tag\NUM.\num$$
\comment
This integral representation follows with the help of [\GRA, p.318]
($\Re(\nu)>-1$):
\plus
$$\int_u^\infty(x-u)^\nu e^{-\mu x}dx
  =\mu^{-\nu-1}\Gamma(\nu+1)e^{-u\mu}.
  \tag\NUM.\num$$
\endcomment
Therefore we get for the $\zeta_\Gamma(c;s)$-contribution [\AOK, \GROd]
in Eq.(\numeh) ($m$ even)
\plus
$$\zeta_\Gamma(s;c)=
  {\sin\pi s\over8\pi}
  \int_0^\infty{d\lambda\over[\lambda(\lambda+2c)]^s}
  {d\over d\lambda}\ln\left[
  {\mZO(\lambda+c+1+{m\over2})\mZO(\lambda+c-{m\over2})\over
   \mZE(\lambda+c+{1-m\over2})\mZE(\lambda+c+{m+1\over2})}\right],
  \tag\NUM.\num$$
where the logarithmic derivative of the super zeta-functions has been
used. Let be $f(s)=\sin(\pi s)[\lambda(\lambda+2c)]^{-s}]$. Then $f'(s)
|_{s=0}=\pi$ and we get for $\zeta^{\prime}(0;c)$ $(\Re(s)>0$):
\plus
$$\zeta'(0;c)={\hatCA\over8\pi}\ln c^2
   -\half\ln\left[{\mZO(c+1+{m\over2})\mZO(c-{m\over2})
   \over\mZE(c+{1-m\over2})\mZE(c+{m+1\over2})}\right].
  \tag\NUM.\num$$
\comment
Here it has been  have used that $\lim_{s\to\infty}{\widehat Z}_q(s)=1$,
which follows at once from the Euler product representation of the
Selberg super zeta-functions.
\endcomment
Performing the limit $c\to\epsilon$ for
$|\epsilon|\ll1$, I get for $m=0$ and $m=-1$, respectively
$$\allowdisplaybreaks\align
  \sdet(-\square_0^2)
  &=\pi^{\hatCA/4\pi}{\mZO(1)\over{\mZE}(\half)}
  \sqrt{\mZE(0)\over\mZE(1)},
  \tag\NUM.\num\\   \global\plus
  \sdet(-\square_{-1}^2)
  &=\pi^{-\hatCA/4\pi}{\mZE(\half)\over{\mZO}(1)}
  \sqrt{\mZE(1)\over\mZE(0)}.
  \tag\NUM.\num\endalign$$
In the general case we obtain
\advance\glno by -2
$$\allowdisplaybreaks\alignat 3
  \sdet(-\square_m^2)
  &=\bigg({\pi\over m!}\bigg)^{\hatCA/4\pi}
    {\mZO(1+{m\over2})\over{\mZE}({m+1\over2})}
  \sqrt{\mZE(0)\over\mZE(1)}, &\qquad &m=2,4,\dots
  \tag\NUM.\num\\    \global\plus
  \sdet(-\square_{-m}^2)
  &=\bigg({(m-2)!\over\pi}\bigg)^{\hatCA/4\pi}
    {\mZO({m\over2})\over{\mZE}({m+1\over2})}
  \sqrt{\mZE(1)\over\mZE(0)}, &\qquad &m=2,4,\dots
  \tag\NUM.\num\\    \global\plus
  \sdet(-\square_m^2)
  &=\bigg({\pi\over im!}\bigg)^{\hatCA/4\pi}
    {\mZE(1+{m\over2})\over{\mZO}({m+1\over2})}
  \sqrt{\mZE(0)\over\mZE(1)}, &\qquad &m=1,3,\dots
  \tag\NUM.\num\\    \global\plus
  \sdet(-\square_{-m}^2)
  &=\bigg({i(m-2)!\over\pi}\bigg)^{\hatCA/4\pi}
    {\mZE({m\over2})\over{\mZO}({m+1\over2})}
  \sqrt{\mZE(1)\over\mZE(0)}, &\qquad &m=3,5,\dots
  \tag\NUM.\num\endalignat$$
(note for instance the relation $\sdet(-\square_0^2)\cdot
\sdet(-\square_{-1}^2)=1$). Here, of course, use has been made of the
functional relations for the modified Selberg super zeta-functions. In
particular we get [c.f.\ Eq.(\numac)]
\plus
$$[\sdet(-\square_0^2)]^{-5/2}[\sdet(-\square_2^2)]^{1/2}
  ={\pi^{-\hatCA/2\pi}\over\sqrt{2}}
   \bigg({\mZO(1)\over\mZE(\half)}\bigg)^{-5/2}
   \bigg({\mZO(2)\over\mZE({3\over2})}\bigg)^{1/2}
   {\mZE(0)\over\mZE(1)},
  \tag\NUM.\num$$
These determinants are the ones for the Dirac-Laplace operator for
Dirichlet boundary-conditions on bordered super Riemann surface. In
order to distinguish from the those with Neumann boundary-conditions,
${\sdet^{(N)\,\prime}_{\Sigma}}(-\square_0^2)$, we denote therefore
$\sdet(-\square_0^2)\equiv\sdet^{(D)}_{\Sigma}(-\square_0^2)$. Now we
know that the Selberg super zeta-functions have concerning the
$\gamma\CI$-length product the reverse power behaviour, denoted by an
index ``$(N)$'', i.e.\ $Z^{(N)}(s)$. Furthermore we have to take into
account that instead of bosonic and fermionic eigenfunctions which are
odd with respect to $x$ of $\square_0$, we have that bosonic and
fermionic eigenfunctions which are even with respect to $x$ appear,
i.e.\ we have for instance
\plus
$${\sdet^{(N)\,\prime}_{\Sigma}}
  (-\square_0^2)=(-1)^{1-2q}\pi^{\hatCA/4\pi}
  {\mZEN(1)\over{\widetilde Z}_1^{(N)}(\half)}
  \sqrt{\mZEN(0)\over\mZEN(1)}.
  \tag\NUM.\num$$
Here by ${\widetilde Z}_1^{(N)}(\half)$ the order of ${\hat Z}_1^{(N)}$
at $s=\half$ is denoted, depending whether $\Delta n_0^{(0)}\leq0$ or
$\Delta n_0^{(0)}>0$, respectively. $\Delta n_0^{(0)}=n_0^B-n_0^F$
denotes the difference between the number of even bosonic- and
fermionic zero-modes of the Dirac-Laplace operator $\square_0$.
According to Ref.[\NINN] $\Delta n_0^{(0)}=1-2q$ with
$q=\dim\ker\bar\partial_1$ and $\bar\partial_p^{\dag}
=-y^2\partial_z+\half py$. From the corresponding expressions for
$\sdet^{(D)}_{\Sigma}(-\square_m^2)$ and ${\sdet^{(N)\,\prime}_{\Sigma}}
(-\square_m^2)$, respectively, now follows
\plus
$$\sdet^{(D)}_{\Sigma}(-\square_m^2)\cdot
  {\sdet^{(N)\,\prime}_{\Sigma}}(-\square_m^2)
  ={\sdet_{\widehat\Sigma}}'(-\square_m^2),
  \tag\NUM.\num$$
\edef\numfa{\NUM.\num}%
where $\widehat\Sigma$ denotes the closed double of the bordered super
Riemann surface. The corresponding Selberg super zeta-functions on
$\widehat\Sigma$ are then defined by [\GROd]
\plus
$$Z_q(s)=\prod_{\{\gamma\}}\prod_{k=0}^\infty
         \Big(1-\chi_\gamma^q e^{-(s+k)l_\gamma}\Big)
  \tag\NUM.\num$$
($q=0,1$). Equation (\numfa) shows in a nice way the super-analogue of
Eq.(\numab) and concludes the discussion.


\bigskip\goodbreak\noindent
\glno=0                
\advance\chapno by 1   
\line{\bf VII.\ Discussion and Summary\hfill}
\par\medskip\nobreak\noindent
In this paper I have discussed a super extension of the Selberg trace
formula for bordered Riemann surfaces, the Selberg supertrace formula
for bordered super Riemann surfaces, including hyperbolic, elliptic and
parabolic conjugacy classes. In the case of the incorporation of
parabolic conjugacy classes an appropriate regularization scheme had to
be applied. In the case of Dirichlet boundary-conditions, Eisenstein
series representing the continuous spectrum of the Dirac-Laplace
operator $\square$ were not needed because the have even parity with
respect to $x$ and they dropped out.

Furthermore, I could discuss Selberg super zeta-functions. Similarly as
in the usual case, there appeared additional ``trivial zeros'' and
``trivial poles'' in comparison to the ``trivial zeros'' of the super
zeta-function due to the additional elliptic and parabolic conjugacy
classes. In particular, the elliptic and parabolic conjugacy classes
altered the multiplicity of the trivial poles already due to the
hyperbolic conjugacy classes, the parabolic terms introduced new
structure.

I also could calculate super-determinants of the Dirac-Laplace operators
on bordered super Riemann surfaces, for Dirichlet and Neumann
boundary-conditions respectively. These determinants were expressed in
terms of the Selberg super-zeta functions which gave a closed
expression for the integrand in the Polyakov path integral for open
fermionic strings. Within this formalism it could also be shown that
the product of these determinants gave the determinants of the
Dirac-Laplace operators on the doubled surface.

However, there are still open problems and questions. For instance, it
is not difficult to derive the corresponding so-called Weyl's and
Huber's laws for the increase in the number of the energy levels and
the norms of the hyperbolic conjugacy classes, respectively. Whereas in
the former case one gets $\#N [\lambda_n^{(B)}-\lambda_n^{(F)}]\propto{\
CA/4\pi}$ $(\lambda\to\infty)$, which gives a Witten-index, in the
latter one finds $\# N(L)\propto e^{\Delta n_0^{(0)}L}/L$
$(L\to\infty)$ form which follows $\Delta n_0^{(0)}>0$.

The case of super automorphic $m$-forms ($m\not=0$) is only available
for hyperbolic conjugacy classes, which are the most important ones due
to their appearance in the evaluation of determinants. In order to set
up the corresponding supertrace formula which include elliptic and
parabolic conjugacy classes as well, the inversion formul\ae\
Eqs.(\numcd,\numce) must be exploited which are very involved and give
rise to considerably complicated expressions.

Finally, we must note that it is not clear for general non-cocompact
(super) Fuchsian groups, whether the Maass-Laplacian $\Delta_m$ and its
super counterpart $\square_m$ have infinitely many eigenvalues. In
fact, for non-arithmetic groups a conjecture by Phillips and Sarnak
[\PHSAR] says that this will not be the case. In the case of arithmetic
groups it is then possible to evaluate all contributions needed for the
calculation of determinants in the trace formula [\KOYA]. However, it
is nevertheless possible to {\it define} (super-) determinants by means
of Selberg (super) zeta-functions which are constructed in the usual
way (see Takhtajan and Zograf [\TAZO] for a discussion). However, a
treatment of this matter for the case of (bordered) super Riemann
surfaces is beyond the scope of this paper and is devoted to future
investigations.

\bigskip\goodbreak\noindent
\line{\bf Acknowledgement\hfill}
\par\nobreak\noindent
Fruitful discussions are gratefully acknowledged with J.\ Bolte,
H.\ Ninnemann, C.\ Reina and F.\ Steiner. Special thanks are given to
D.\ Hejhal, and I thank K.\ Aoki and S.\ B.\ Giddings for helpful
communications. I would also like to thank the members of the
II.Institut f\"ur Theoretische Physik, Hamburg University, for their
kind hospitality.

\bigskip\goodbreak\noindent
\line{\bf References\hfill}
\par\nobreak\noindent
\eightpoint
\eightrm
\baselineskip=11pt
\item{[\ALV]}
Alvarez, O.:
Theory of Strings with Boundaries:
Fluctuations, Topology and Quantum Theory.
Nucl.Phys.\ {\bf B 216}, 125-184 (1983)
\item{[\ALVA]}
Alvarez-Gaum\'e, L., Moore, G.\ and Vafa, C.:
Theta Functions, Modular Invariance, and Strings.
Commun.Math.Phys.\ {\bf 106}, 1-40 (1986)
\item{[\AOK]}
Aoki, K.:
Heat Kernels and Super Determinants of Laplace Operators on Super
Riemann Surfaces.
Commun.Math.Phys.\ {\bf 117}, 405-429 (1988)
\item{[\BMFSa]}
Baranov, A.M., Manin, Yu.I., Frolov, I.V.\ and Schwarz, A.S.:
The Multiloop Contribution in the Fermionic String.
Sov.J.Nucl.Phys.\ {\bf 43}, 670-671 (1986)
\item{[\BMFSb]}
Baranov, A.M., Manin, Yu.I., Frolov, I.V.\ and Schwarz, A.S.:
A Superanalog of the Selberg Trace Formula and Multiloop
Contributions for Fermionic Strings.
Commun.Math.Phys.\ {\bf 111}, 373-392 (1987)
\item{[\BFS]}
Baranov, A.M., Frolov, I.V.\ and Shvarts, A.S.:
Geometry of Two-Dimensional Superconformal Field Theories.
Theor.Math.Phys.\ {\bf 70}, 64-72 (1987)
\item{[\BASCH]}
Baranov, A.M.\ and A.S.Schwarz, A.S.:
Multiloop Contribution to String Theory.
JETP Lett.\ {\bf 42}, 419-421 (1985);
On the Multiloop Contributions to the String Theory.
Int.J.Mod.Phys.\ {\bf A 2}, 1773-1796 (1987)
\item{[\BATCH]}
Batchelor, M.:
The Structure of Supermanifolds.
Trans.Amer.Math.Soc.\ {\bf 253}, 329-338 (1979);
Two Approaches to Supermanifolds.
Trans.Amer.Math.Soc.\ {\bf 258}, 257-270 (1980)
\item{[\BABR]}
Batchelor, M.\ and Bryant, P.:
Graded Riemann Surfaces.
Commun.Math.Phys.\ {\bf 114}, 243-255 (1988)
\item{[\BLCL]}
Blau, S.K.\ and Clements, M.:
Determinants of Laplacians for World Sheets with Boundaries.
Nucl. Phys.\ {\bf B 284}, 118-130 (1987)
\item{[\BCDPC]}
Blau, S.K., Clements, M., Della Pietra, S., Carlip, S.\ and
Della Pietra, V.:
The String Amplitude on Surfaces with Boundaries and Crosscaps.
Nucl.Phys.\ {\bf B 301}, 285-303 (1988)
\item{[\BOGRO]}
Bolte, J.\ and Grosche, C.:
Selberg Trace Formula for Bordered Riemann Surfaces: Hyperbolic,
Elliptic and Parabolic Conjugacy Classes, and Determinants of
Maass-Laplacians; {\it DESY-Pre\-print} DESY 92 - 118, and
{\it SISSA-Pre\-print}, SISSA/\-139/\-92/\-FM, August 1992
\item{[\BOSTa]}
Bolte, J.\ and Steiner, F.:
Determinants of Laplace-Like Operators on Riemann Surfaces.
Commun. Math.Phys.\ {\bf 130}, 581-597 (1990)
\item{[\BOSTb]}
Bolte, J.\ and Steiner, F.:
The Selberg Trace Formula for Bordered Riemann Surfaces.
{\it DESY Preprint}, DESY 90-082, 1-14, July 1990
\item{[\BUMO]}
Burgess, C.\ and Morris, T.R.:
Open and Unoriented Strings \`a la Polyakov.
Nucl.Phys.\ {\bf B 291}, 256-284 (1987);
Open Superstrings \`a la Polyakov.
Nucl.Phys.\ {\bf B 291}, 285-333 (1987)
\item{[\CARL]}
S.Carlip:
Sewing Closed String Amplitudes.
Phys.Lett.\ {\bf B 209} 464-472 (1988)
\item{[\DEBRO]}
De Beer, W.\ and Rodriguez, J.P.:
Holomorphic Factorization and Open Strings.
Phys.Lett.\ {\bf B 213}, 291-297 (1988)
\item{[\DEW]}
DeWitt, B.:
Supermanifolds.
Cambridge: Cambridge University Press, 1984
\item{[\DHPHa]}
D'Hoker, E.\ and Phong, D.H.:
Loop Amplitudes for the Bosonic Polyakov String.
Nucl.Phys.\ {\bf B 269}, 205-234 (1986)
\item{[\DHPHb]}
D'Hoker, E.\ and Phong, D.H.:
On Determinants of Laplacians on Riemann Surfaces.
Commun.Math. Phys.\ {\bf 104}, 537-545 (1986)
\item{[\DHPHc]}
D'Hoker, E.\ and Phong, D.H.:
Loop Amplitudes for the Fermionic String.
Nucl.Phys.\ {\bf B 278}, 225-241 (1986)
\item{[\DHPHd]}
D'Hoker, E.\ and Phong, D.H.:
The Geometry of String Perturbation Theory.
Rev.Mod.Phys.\ {\bf 60}, 917-1065 (1988)
\item{[\DUN]}
Dunbar, D.C.:
Boundary and Crosscap States in Compactification of Open and Closed
Bosonic Strings.
Int.J.Mod.Phys.\ {\bf A 4}, 5149-5176 (1989)
\item{[\EFR]}
Efrat, I.:
Determinants of Laplacians on Surfaces of Finite Volume.
Commun.Math.Phys.\ {\bf 119}, 443-451 (1988); and
Erratum: Commun.Math.Phys.\ {\bf 119}, 607 (1991)
\item{[\ELS]}
Elstrodt, J.:
Die Resolvente zum Eigenwertproblem der automorphen Formen in der
hyperbolischen Ebene. Teil I.
Math.Ann.\ {\bf 203}, 295-330 (1973)
\item{[\EMOTa]}
Erd\'elyi, A., Magnus, W., Oberhettinger, F., and Tricomi, F.G. (eds.):
Higher Transcendental Functions, Vol.I.
New York: McGraw Hill, 1954
\item{[\GIL]}
Gilbert, G.:
String Theory Path Integral - Genus Two and Higher.
Nucl.Phys.\ {\bf B 277}, 102-124 (1986)
\item{[\GRA]}
Gradshteyn, I.S.\ and Ryzhik, I.M.:
Table of Integrals, Series, and Products.
New York: Academic Press, 1980
\item{[\GRE]}
Green, M.B.:
Supersymmetrical Dual String Theories and Their Field Theory Limits.
{Surveys in High Energy Physics} {\bf 3}, 127-160 (1983)
\item{[\GSa]}
Green, M.B.\ and Schwarz, J.H.:
Supersymmetrical Dual String Theory, I-III.
Nucl. Phys.\ {\bf B 181}, 502-530 (1982), {\bf B 198}, 252-268,
441-460 (1982)
\item{[\GSb]}
Green, M.B.\ and Schwarz, J.H.:
Anomaly Cancelations in Supersymmetric D=10 Gauge Theory and
Superstring Theory.
Phys.Lett.\ {\bf B 149}, 117-122 (1984);
Infinity Cancelations in SO(32) Superstring Theory.
Phys.Lett.\ {\bf B 151}, 21-25 (1985);
The Hexagon Gauge Anomaly in Type I Superstring Theory.
Nucl.Phys.\ {\bf B 255}, 93-114 (1985)
\item{[\GSW]}
Green, M.B., Schwarz, J.H.\ and Witten, E.:
Superstring Theory I\&II.
Cambridge: Cambridge University Press, 1988
\item{[\GROd]}
Grosche, C.:
Selberg Supertrace Formula for Super Riemann Surfaces, Analytic
Properties of the Selberg super zeta-functions and Multiloop
Contributions to the Fermionic String.
{\it DESY Preprint} DESY 89 - 010, February 1989, 1-102,
and Commun.Math.Phys.\ {\bf 133}, 433-485 (1990)
\item{[\GROe]}
Grosche, C.:
Selberg Supertrace Formula for Super Riemann Surfaces II:
Elliptic and Parabolic Conjugacy Classes,
and Selberg Super Zeta-Functions.
Commun.Math.Phys.\ {\bf 151}, 1-37 (1993)
\item{[\GROf]}
Grosche, C.:
Selberg Trace-Formul\ae\ in Mathematical Physics;
{\it SISSA-Pre\-print}, SISSA/\-177/\-92/\-FM, October 1992,
to appear in: {\it Proceedings of the Workshop ``From Classical to
Quantum Chaos (1892-1992)'', 21-24 July, 1992, Trieste}, World
Scientific, Singapore; eds.: G.\ Dell'Antonio, S.\ Fantoni and V.\ R.\
Manfredi.
\item{[\GP]}
Gross, D.J.\ and Periwal, V.:
String Perturbation Theory Diverges.
Phys.Rev.Lett.\ {\bf 60}, 2105-2108 (1988)
\item{[\HEJ]}
Hejhal, D.A.:
The Selberg Trace Formula and the Riemann Zeta Function.
Duke Math.J.\ {\bf 43}, 441-482 (1976)
\item{[\HEJa]}
Hejhal, D.A.:
The Selberg Trace Formula for $\PSL(2,\bbbr)$, I,
Lecture Notes in Mathematics, Vol.548.
Berlin, Heidelberg, New York: Springer, 1976
\item{[\HEJb]}
Hejhal, D.A.:
The Selberg Trace Formula for $\PSL(2,\bbbr)$, II,
Lecture Notes in Mathematics, Vol.1001.
Berlin, Heidelberg, New York: Springer, 1981
\item{[\HOWE]}
Howe, P.S.:
Superspace and the Spinning String.
Phys.Lett.\ {\bf B 70}, 453-456 (1977);
Super Weyl Transformations in Two Dimensions.
J.Phys.A: Math.Gen.\ {\bf 12}, 393-402 (1979)
\item{[\JASK]}
Jaskolski, Z.:
The Polyakov Path Integral over Bordered Surfaces (the Open String
Amplitudes).
Commun.Math.Phys.\ {\bf 128}, 285-318 (1990);
The Polyakov Path Integral over Bordered Surfaces (the Closed String
Off-Shell Amplitudes).
Commun.Math.Phys.\ {\bf 139}, 353-376 (1991)
\item{[\KOYA]}
Koyama, S.-Y.:
Determinant Expression of Selberg Zeta Functions. I-III.
Trans.Amer.Math.Soc.\ {\bf 324}, 149-168 (1991);
Trans.Amer.Math.Soc.\ {\bf 329}, 755-772 (1992);
Proc.Amer.Math.Soc.\ {\bf 113}, 303-311 (1991)
\item{[\LOSEV]}
Losev, A.S.:
Calculation of a Scalar Determinant in the Theory of Open Strings.
JETP Lett.\ {\bf 48}, 330-333 (1988)
\item{[\LUCK]}
Luckock, H.:
Quantum Geometry of Strings with Boundaries.
Ann.Phys.(N.Y.)\ {\bf 194}, 113-147 (1989)
\item{[\MANIN]}
Manin, Yu.I.:
The Partition Function of the Polyakov String can be Expressed in Terms
of Theta Functions.
Phys.Lett.\ {\bf B 172}, 184-185 (1986)
\item{[\MDM]}
Mart\'\i n-Delgado, M.A.\ and Mittelbrunn, J.R.:
Bordered Riemann Surfaces, Schottky Groups and Off-Shell String
Amplitudes.
Int.J.Mod.Phys.\ {\bf A 6}, 1719-1747 (1991)
\item{[\MUYa]}
Matsumoto, S., Uehara, S.\ and Yasui, Y.:
Hadamard Model on the Super Riemann Surface.
Phys. Lett.\ {\bf A 134}, 81-86, (1988)
\item{[\MUYb]}
Matsumoto, S., Uehara, S.\ and Yasui, Y.:
A Superparticle on the Super Riemann Surface.
J.Math. Phys.\ {\bf 31}, 476-501 (1990)
\item{[\MNP]}
Moore, G., Nelson, P.\ and Polchinski, J.:
Strings and Supermoduli.
Phys.Lett.\ {\bf B 169}, 47-53 (1986)
\item{[\MORO]}
Mozorov, A.\ and Rosly, A.:
On Many-Loop Calculations in the Theory of Open Strings.
Phys.Lett.\ {\bf B 214}, 522-526 (1988);
Some Examples of Computation of the Scalar Determinant in Open
String Theory.
Theor.Math.Phys.\ {\bf 80}, 899-911 (1989);
and Nucl.Phys.\ {\bf B 326}, 185-204 (1989);
Strings and Riemann Surfaces;
Nucl.Phys.\ {\bf B 326}, 205-221 (1989)
\item{[\NARA]}
Namazie, M.A.\ and Rajjev, S.:
On Multiloop Computations in Polyakov's String Theory.
Nucl.Phys.\ {\bf B 277}, 332-348 (1986)
\item{[\NINN]}
Ninnemann, H.:
Holomorphe and Harmonische Formen auf Super-Riemannschen Fl\"achen und
ihre Anwendung auf den Fermionischen String.
{\it Diploma Thesis}, Hamburg University, 1989;
\newline
Deformations of Super Riemann Surfaces.
{\it Commun.Math.Phys.}\ {\bf 150}, 267--288 (1992)
\item{[\OHN]}
Ohndorf, T.:
Covariant Operator Formalism of the Bosonic String and Polyakov's
Path Integral on Bordered Riemann Surfaces.
{\it Heidelberg Preprint}, HD-THEP-1988-37, 1-140, October 1988
\item{[\OSH]}
Oshima,K.:
Completeness Relations for Maass Laplacians and Heat Kernels on the
Super Poincar\'e Upper Half-Plane.
J.Math.Phys.\ {\bf 31}, 3060-3063 (1990)
\item{[\PHSAR]}
Phillips, R.S.\ and Sarnak, P.:
On Cusp Forms for Cofinite Subgroups of $\PSL(2,\bbbr)$.
Invent.Math.\ {\bf 80}, 339-364 (1985)
\item{[\POLa]}
Polyakov, A.M.:
Quantum Geometry of Bosonic Strings.
Phys.Lett.\ {\bf B 103}, 207-210 (1981)
\item{[\POLb]}
Polyakov, A.M.:
Quantum Geometry of Fermionic Strings.
Phys.Lett.\ {\bf B 103}, 211-213 (1981)
\item{[\RC]}
Rabin, J.M.\ and Crane, L.:
Global Properties of Supermanifolds.
{Commun.Math. Phys]} {\bf 100}, 141-160 (1985);
How Different are the Supermanifolds of Rogers and DeWitt?.
Commun.Math.Phys.\ {\bf 102}, 123-137 (1985);
Super Riemann Surfaces: Uniformization and Teichm\"uller Theory.
Commun.Math. Phys.\ {\bf 113}, 601-623 (1988)
\item{[\RODR]}
Rodriguez, J.P.:
Open Strings from Closed Strings: Period Matrix, Measure and Ghost
Determinant.
Phys.Lett.\ {\bf B 202}, 227-232 (1988)
\item{[\ROVT]}
Rodriguez, J.P.\ and Van Tander, A.:
Spin Structures for Riemann Surfaces with Boundaries and Cross-Caps.
Phys.Lett.\ {\bf B 217}, 85-90 (1989)
\item{[\ROG]}
Rogers, A.:
A Global Theory of Supermanifolds.
J.Math.Phys.\ {\bf 21}, 1352-1364 (1980);
On the Existence of Global Integral Forms on Supermanifolds.
J.Math.Phys.\ {\bf 26}, 2749-2753 (1985);
Graded Manifolds, Supermanifolds and Infinite-Dimensional Grassmann
Algebras.
Commun.Math.Phys.\ {\bf 105}, 375-384 (1986)
\item{[\SCHW]}
Schwarz, J.H.:
Superstring Theory.
Phys.Rep.\ {\bf 89}, 223-322 (1982)
\item{[\SELB]}
Selberg, A.:
Harmonic Analysis and Discontinuous Groups in Weakly Symmetric
Riemannian Spaces With Application to Dirichlet Series.
J.Indian Math.Soc.\ {\bf 20}, 47-87 (1956)
\item{[\SHOK]}
Shohkranian, S.:
The Selberg-Arthur Trace Formula,
Lecture Notes in Mathematics, Vol.1503.
Berlin, Heidelberg, New York: Springer, 1992
\item{[\SIBN]}
Sibner, R.J.:
Symmetric Fuchsian Groups.
Am.J.Math.\ {\bf 90}, 1237-1259 (1968)
\item{[\STEI]}
Steiner, F.:
On Selberg's Zeta Function for Compact Riemann Surfaces.
Phys.Lett.\ {\bf B 188}, 447-454 (1987);
Quantum Chaos and Geometry. In:
Mitter, H.\ and Pittner, L.(eds.):
Recent Developments in Mathematical Physics,
26.Internationale Universit\"atswochen, Schladming 1987, 305-312.
Berlin, Heidelberg, New York: Springer, 1987
\item{[\TAZO]}
Takhtajan, L.A.~and Zograf, P.G.:
A Local Index Theorem for Families of $\bar\partial$-Operators on
Punctured Riemann Surfaces and a New K\"ahler Metric on Their Moduli
Spaces.
{Commun.Math.Phys]}~{\bf 137}, 399-426 (1991)
\item{[\UEYA]}
Uehara, S.\ and Yasui, Y.:
A Superparticle on the ``Super'' Poincar\'e Upper Half Plane.
Phys.Lett.\ {\bf B 202}, 530-534 (1988);
Super-Selberg's Trace Formula from the Chaotic Model.
J.Math.Phys\ {\bf 29}, 2486-2490 (1988)
\item{[\VENa]}
Venkov, A.B.:
On an Asymptotic Formula Connected with the Number of Eigenvalues
Corresponding to Odd Eigenfunctions of the Laplace-Beltrami Operator
on a Fundamental Region of the Modular Group $\PSL(2,\bbbz)$.
Sov.Math.Dokl.\ {\bf 18}, 524-526 (1977);
Selberg's Trace Formula and Non-Euclidean Vibrations of an Infinite
Membrane.
Sov.Math.Dokl.\ {\bf 19}, 708-712 (1979);
Selberg's Trace Formula for the Hecke Operator Generated by an
Involution, and the eigenvalues of the Laplace-Beltrami Operator on the
Fundamental Domain of the Modular Group $\PSL(2,\bbbz)$.
Math.USSR.Izv.\ {\bf 12}, 448-462 (1978)
\item{[\VENb]}
Venkov, A.B.:
Spectral Theory of Automorphic Functions, the Selberg Trace Formula,
and Some Problems of Analytic Number Theory and Mathematical Physics.
Russian Math.Surveys {\bf 34}, 79-153 (1979)
\item{[\VENc]}
Venkov, A.B.:
Spectral Theory of Automorphic Functions.
Proc.Math.Inst.Steklov {\bf 153}, 1-163 (1981)
\item{[\VENd]}
Venkov, A.B.:
Spectral Theory of Automorphic Functions and Its Applications.
Dordrecht: Kluwer Academic Publishers, 1990
\item{[\VOROS]}
Voros, A.:
The Hadamard Factorization of the Selberg Zeta Function on a Compact
Riemann Surface.
Phys.Lett.\ {\bf B 180}, 245-246 (1986);
\newline
Spectral Functions, Special Functions and the Selberg Zeta Function.
Commun.Math.Phys.\ {\bf 110}, 439-465 (1987)
\item{[\WU]}
Wu, S.:
Determinants of Dirac Operators and Thirring Model Partition Functions
on Riemann Surfaces with Boundaries.
Commun.Math.Phys.\ {\bf 124}, 133-152 (1989)


\enddocument